\documentclass[onecolumn,draftcls]{IEEEtran}

\usepackage{amsmath,amsfonts,amssymb}
\usepackage{graphicx,color}
\usepackage{pgfplots} 
\pgfplotsset{compat=1.14}
\usepackage{stackrel,cite}
\usepackage{braket,notation}

\newcommand{\refE}[1]   {(\ref{eqn:#1})}
\newcommand{\refS}[1]   {Section \ref{sec:#1}}

\newcommand{\Sphi}[1]{\Ec^{\bullet}_{#1}}
\newcommand{\Spho}[1]{\Ec^{\circ}_{#1}}

\def\BibTeX{{\rm B\kern-.05em{\sc i\kern-.025em b}\kern-.08em
    T\kern-.1667em\lower.7ex\hbox{E}\kern-.125emX}}
\begin{document}
\title{Generalized Perfect Codes for\\Symmetric Classical-Quantum Channels}
\author{Andreu~Blasco~Coll,~\IEEEmembership{Student~Member,~IEEE,}
        Gonzalo~Vazquez-Vilar,~\IEEEmembership{Member,~IEEE,}
        Javier~R.~Fonollosa,~\IEEEmembership{Senior~Member,~IEEE}
\thanks{G. Vazquez-Vilar is also with the Gregorio Mara\~n\'on Health Research Institute, Madrid, Spain. This work has been funded by the ERC grant 714161, by the AEI of Ministerio de Ciencia, Innovaci\'on y Universidades of Spain, \mbox{TEC2016-75067-C4-2-R}, \mbox{TEC2016-78434-C3-3-R}, \mbox{RED2018-102668-T} and \mbox{PID2019-104958RB-C41} with ESF and Dept. d'Empresa i Coneixement de la Generalitat de Catalunya, 2017 SGR 578 AGAUR and 001-P-001644 QuantumCAT within the ERDF Program of Catalunya.}
\thanks{This work was presented in part at the 2016 IEEE International Symposium on Information Theory, Barcelona, Spain.}}

\maketitle

\begin{abstract}
We define a new family of codes for symmetric classical-quantum channels and establish their optimality. To this end, we extend the classical notion of generalized perfect and quasi-perfect codes to channels defined over some finite dimensional complex Hilbert output space. The resulting optimality conditions depend on the channel considered and on an auxiliary state defined on the output space of the channel. For certain $N$-qubit classical-quantum channels, we show that codes based on a generalization of Bell states are quasi-perfect and, therefore, they feature the smallest error probability among all codes of the same blocklength and cardinality.
\end{abstract}


\section{Introduction}
In the context of reliable communication, the ultimate goal of information theory is to characterize the best achievable performance of any transmission scheme and to establish the structure of codes and decoders attaining this limit. While information theory has not reached this goal in general, in certain regimes the best performance of a communication system is accurately characterized and there exist practical codes attaining  it.
In his landmark paper \cite{Shan48}, Shannon demonstrated that for every communication channel there exists a fundamental limit, named channel capacity, that determines the highest rate at which a sender can transmit data to a receiver with arbitrarily small decoding error probability, provided that we employ a sufficiently long error correcting code. Nowadays, several code constructions achieve the channel capacity or perform very close to it. Specific examples, widely used in current communication systems, are low-density parity check (LDPC) codes\cite{urbankerichardson08}, turbo codes \cite{berrouglavieuxthitimajshima93}, or polar codes \cite{arikan09}. 
If the length of the code is limited --e.g., due to delay constraints or due to the nature of the channel-- the channel capacity is not a good benchmark anymore. To accurately describe the performance limits of the system in this regime, we shall use non-asymptotic bounds on the error probability (or on the transmission rate) of the best coding scheme. Two instances of these limits are the sphere-packing bound~\cite[Eq. (5.8.19)]{Gall68} for the binary symmetric channel (BSC) and Shannon's non-asymptotic results for the additive white Gaussian noise (AWGN) channel~\cite{shannon59gaussian}. For general channels, Polyanskiy \textit{et al.}  proposed several (upper and lower) bounds which accurately characterize the performance of communication systems in the finite blocklength regime \cite{Pol10}. Code designers have optimized the finite-length performance of certain codes and now they perform close to those non-asymptotic limits (see, e.g., \cite{coskun2019,bocharova20} and references therein). 
Moreover, certain codes can even attain these limits with equality, thus proving their non-asymptotic optimality. For example, perfect and quasi-perfect binary codes attain the sphere-packing bound~\cite[Eq. (5.8.19)]{Gall68} for the BSC. 
The notion of pefect and quasi-perfect codes was generalized beyond binary alphabets in \cite{tit19qp}.
These codes, whenever they exist, attain the hypothesis-testing bound \cite[Th.~27]{Pol10} with equality
and include, e.g., maximum-distance separable (MDS) codes for erasure channels. 


The results presented above consider transmission channels --or random transformations-- which are modeled by a transition probability distribution. Certain physical systems, however, can only be described using the laws of quantum mechanics. For these systems, the classical channel capacity and the corresponding non-asymptotic results
have to be extended to encompass the quantum properties of the system. Holevo, Schumacher and Westmoreland studied the task of sending classical data over a channel with classical inputs and quantum outputs~\cite{holevo98,schumacher97}; this setting is usually referred to as \textit{classical-quantum channel coding}. Their coding theorem guarantees the existence of reliable codes if their rate is below a fundamental limit, known as \textit{Holevo capacity}, provided that the codelength is sufficiently long. While the proof of this result does not provide an explicit code construction, it guided the design of practical coding schemes. For example, quantum polar codes are practical constructions shown to attain this limit \cite{wilde13}, the codes proposed in~\cite{sasaki97,sasaki98} feature the superadditivity of mutual information present in Holevo capacity, and other coding schemes exploiting the quantum propertiesof optical channels were proposed in \cite{guha11, dimario19}. 
Holevo capacity is an asymptotic quantity that, in general, can only be attained by a large number of channel uses via a joint measurement on the combined channel outputs. For a finite number of channel uses --which is relevant for practical quantum systems-- non-asymptotic performance limits need to be used. Converse non-asymptotic bounds were studied in \cite{hayashi2003general}, \cite[Sec. 4.6]{Hayashi06} and \cite{matthews2014finite}, among other works. However, to the best of our knowledge, these works have not been applied in the design and/or benchmark of practical code constructions.

A separate line of research considers the transmission of quantum information in a noisy environment. In \cite{shor95}, Shor showed that quantum errors can be controlled by encoding the state of the system in a quantum code and periodically performing measurements on the redundant parts of the code. This observation opened the field of \textit{quantum error correction}. While it is possible to encapsulate classical information over noisy quantum channels using quantum error correction codes, they are highly inefficient for this task. We may conclude that, while there is some incipient ongoing work, much less is known about the structure of optimal codes for classical-quantum channels compared to the classical setting.

In this work, we study the structure of optimal codes for certain classical-quantum channels and their connection with quantum hypothesis testing. In particular, we derive two alternative expressions for the error probability of quantum multiple hypothesis testing, which are then used to determine the exact error probability for a fixed classical-quantum channel code. A weakening of this result yields the non-asymptotic converse bound \cite[Eq.~(45)]{matthews2014finite} (see also \cite[Sec. 4.6]{Hayashi06}), hence providing a tool for better understanding previous results in the literature.
We introduce a new family of codes that extends the notion of generalized perfect and quasi-perfect codes~\cite{tit19qp} to symmetric classical-quantum channels.
These codes, whenever they exist, are shown to attain the converse bound with equality and, henceforth, they are optimal. While these codes are possibly rare, we characterize a family of codes based on the Bell states which are quasi-perfect for certain non-asymptotic 2-qubit classical-quantum channels and their $N$-qubit extension.

The organization of this article is as follows. In \refS{QHT} we formalize the problems of binary and multiple hypothesis testing and establish a connection between them. \refS{cq-channels} presents the classical-quantum channel model and establishes the accuracy of different converse bounds in the literature. \refS{qp-codes} defines perfect and quasi-perfect codes for classical-quantum symmetric channels and proves their optimality whenever they exist. In \refS{bell-codes} we study a family of codes which are quasi-perfect for 2-qubit classical-quantum channels affected by quantum erasures or by depolarization. \refS{discussion} concludes the article with some final remarks.

\subsection{Notation}
In the general case, a quantum state is described by a density operator $\rho$ acting on some finite dimensional complex Hilbert space~$\Hc$. Density operators are self-adjoint, positive semidefinite, and have unit trace.
A measurement on a quantum system is a mapping from the state of the system $\rho$ to a classical outcome $m\in\{1,\ldots,M \}$. A measurement is represented by a collection of positive self-adjoint operators $\bigl\{\Pi_1,\ldots,\Pi_M\bigr\}$ such that $\sum \Pi_m = \openone$, where $\openone$ is the identity operator. These operators form a \textit{positive operator-valued measure} (POVM). A POVM measurement $\bigl\{\Pi_1,\ldots,\Pi_M\bigr\}$ applied to $\rho$ has outcome $m$ with probability $\tr(\rho\Pi_m)$ where $\tr$ is the trace operator. 

For self-adjoint operators $A,B$, the notation $A \geq B$ means that $A - B$ is positive semidefinite. Similarly $A \leq B$, $A > B$, and $A < B$ means that $A - B$ is negative semidefinite, positive definite and negative definite, respectively.

For a self-adjoint operator $A$ with spectral decomposition $A = \sum_i \lambda_i E_i$, where $\{\lambda_i\}$ are the eigenvalues and $\{E_i\}$ are the orthogonal projections onto the corresponding eigenspaces, we define
\begin{align}
  \{A > 0\} \triangleq \sum_{i:\lambda_i > 0} E_i.
\end{align}
This corresponds to the projector associated to the positive eigenspace of $A$. We shall also use $\{A \geq 0\} \triangleq \sum_{i:\lambda_i \geq 0} E_i$, $\{A < 0\} \triangleq \sum_{i:\lambda_i < 0} E_i$ and $\{A \leq 0\} \triangleq \sum_{i:\lambda_i \leq 0} E_i$.


\section{Quantum Hypothesis Testing}\label{sec:QHT}

\subsection{Binary Hypothesis Testing}\label{sec:bary}

Let us consider a binary hypothesis test (with simple hypotheses) discriminating between the density operators $\rho_0$ and $\rho_1$ acting on $\Hc$. In order to distinguish between the two hypotheses we perform a measurement. We define a test measurement $\{T, \bar T\}$, such that $T$ and $\bar T \triangleq \openone - T$ are positive semidefinite, self-adjoint operators. The test decides $\rho_0$ (resp. $\rho_1$) when the measurement outcome corresponding to $T$ (resp. $\bar T$) occurs.

Let $\eps_{j|i}$ denote the probability of deciding $\rho_j$ when $\rho_i$ is the true hypothesis, $i,j=0,1$, $i\neq j$. More precisely, 
\begin{align}
 \eps_{1|0}(T) &\triangleq  1- \tr\left(\rho_0 T\right) = \tr\left(\rho_0\bar T\right), \label{eqn:bht-pi10}\\
 \eps_{0|1}(T) &\triangleq \tr\left(\rho_1 T\right).     \label{eqn:bht-pi01}
\end{align}

Let $\alpha_{\beta}(\rho_0 \| \rho_1)$ denote the minimum error probability $\eps_{1|0}$ among all tests with $\eps_{0|1}$ at most $\beta$, that is,
\begin{align}
\alpha_{\beta}(\rho_0 \| \rho_1)
  \triangleq \inf_{T: \eps_{0|1}(T) \leq \beta}  \eps_{1|0}(T).
 \label{eqn:bht-alpha}
\end{align}
The function $\alpha_{\beta}(\cdot\|\cdot)$ is the inverse of the function $\beta_{\alpha}(\cdot\|\cdot)$ appearing in~\cite{matthews2014finite}, which is itself related to the hypothesis-testing relative entropy as $D_{\text{H}}^{\alpha}(\rho_0 \| \rho_1) = - \log \beta_{\alpha}(\rho_0 \| \rho_1)$~\cite{wangrenner2012}.

When $\rho_0$ and $\rho_1$ commute, the test $T$ in~\refE{bht-alpha} can be restricted to be diagonal in the (common) eigenbasis of $\rho_0$ and $\rho_1$, then \refE{bht-alpha} reduces to the classical case~\cite{arxiv15ht}.

The form of the test  minimizing \refE{bht-alpha} is given by the quantum Neyman-Pearson lemma, presented next.

\begin{lemma}[Neyman-Pearson lemma]\label{lem:NPlemma}
The best trade-off between type-I and type-II error probabilities is attained by tests of the form
\begin{align}
  T_{\text{NP}} = \bigl\{\rho_0 - t \rho_1 > 0\bigr\} + \theta^{0}_{t},
  \label{eqn:NPlemma}
\end{align}
for some $t$ and $\theta^{0}_{t}$, and
where $0 \leq \theta^{0}_{t} \leq \bigl\{\rho_0 - t \rho_1 = 0\bigr\}$.
\end{lemma}
\begin{IEEEproof}
A slightly weaker formulation of this result is usually given in the literature (see, e.g., \cite[Ch.~IV, Eq.~(2.18)]{Helstrom76}). The precise statement included here can be found, e.g., in \cite[Lem. 3]{jenvcova2010quantum}. 
\end{IEEEproof}

Then, for any choice of $t$ and $\theta^{0}$ such that $\tr\left(\rho_1 T_{\text{NP}}\right) = \beta$, the resulting test $T_{\text{NP}}$ in \refE{NPlemma} minimizes~\refE{bht-alpha}. 
The following result is a corollary to the Neyman-Pearson lemma that will be useful in the sequel.

\begin{lemma}\label{lem:NPcorollary}
For any binary hypothesis test discriminating between the quantum states $\rho_0$ and $\rho_1$, it follows that
\begin{align}
\!\alpha_{\beta}(\rho_0 \| \rho_1)
&= \sup_{t \geq 0} \Bigl\{ \tr\bigl( \rho_0 \bigl\{\rho_0 - t \rho_1 \leq 0\bigr\} \bigr)
 + t\bigl(\tr\bigl(\rho_1 \bigl\{\rho_0 - t \rho_1 > 0\bigr\}\bigr) \!-\! \beta\bigr) \Bigr\}
\label{eqn:NPcorollary-1}\\
&\geq \tr\bigl( \rho_0 \bigl\{\rho_0 - t' \rho_1 \leq 0\bigr\} \bigr)  - t' \beta,\label{eqn:NPcorollary-2}
\end{align}
for any $t' \geq 0$.
\end{lemma}
\begin{IEEEproof}
The identity \refE{NPcorollary-1} is the quantum analogue of \cite[Lem.~1]{tit19qp} and the relaxation 
\refE{NPcorollary-2} coincides with \cite[Lem.~2]{isit16quantum}.
For completeness, we include next the proof of \refE{NPcorollary-1}-\refE{NPcorollary-2}.

For any operator $A\geq 0$  and $0 \leq T \leq \openone$, it holds that $\tr\bigl( A \{ A > 0\} \bigr) \geq \tr\bigl( A T \bigr)$~\cite[Eq. 8]{nagaoka2007ht}. For $A = \rho_0-t'\rho_1$ and $T=T_{\text{NP}}$ defined in \refE{NPlemma}, this inequality becomes
\begin{align}
  \tr\bigl( (\rho_0-t'\rho_1) P^{+}_{t'} \bigr)
  \geq \tr\bigl( (\rho_0-t'\rho_1) T_{\text{NP}} \bigr),
  \label{eqn:NPbound-1}
\end{align}
where we defined $P^{+}_{t'} \triangleq \{ \rho_0-t'\rho_1 > 0\}$.
Indeed, \refE{NPbound-1} holds with equality for the value $t' = t$
appearing in \refE{NPlemma}, as $\tr\bigl( (\rho_0-t\rho_1) \theta^{0}_{t} \bigr) = 0$ for any $0 \leq \theta^{0}_{t} \leq \bigl\{\rho_0 - t \rho_1 = 0\bigr\}$, tantamount to $\theta^{0}_{t}$ being in the null-space of $\rho_0 - t \rho_1$.

After some algebra, \refE{NPbound-1} yields
\begin{align}
  -\!\tr\bigl( \rho_0 T_{\text{NP}} \bigr)
       \geq -\!\tr\bigl( \rho_0 P^{+}_{t'} \bigr)
            + t' \tr\bigl(\rho_1 (P^{+}_{t'}-T_{\text{NP}}) \bigr).\label{eqn:NPbound-2}
\end{align}
Summing one to both sides of \refE{NPbound-2} and noting that $\alpha_{\beta}(\rho_0 \| \rho_1) = 1-\tr\bigl( \rho_0 T_{\text{NP}}\bigr)$ and $\beta = \tr\bigl(\rho_1 T_{\text{NP}} \bigr)$, we obtain
\begin{align}
\alpha_{\beta}(\rho_0 \| \rho_1)
   \geq \tr\bigl( \rho_0 \{ \rho_0-t'\rho_1 \leq 0\} \bigr)
           \!+t'\!\tr\bigl(\rho_1 P^{+}_{t'}\bigr)\!-t'\beta.\label{eqn:NPbound-3}
\end{align}
As \refE{NPbound-1} holds with equality for the value $t' = t$ appearing in \refE{NPlemma}, so it does \refE{NPbound-3} after optimization over the parameter $t' \geq 0$. Then, \refE{NPcorollary-1} follows.
To obtain the lower bound \refE{NPcorollary-2}, we fix $t'\geq 0$ and use that $\tr\bigl(\rho_1 \bigl\{\rho_0 - t' \rho_1 > 0\bigr\}\bigr)\geq 0$.
\end{IEEEproof}

\subsection{Bayesian Multiple Hypothesis Testing}\label{sec:mary}

We consider now a hypothesis testing problem discriminating among $M$ possible states acting on $\Hc$, where $M$ is assumed to be finite. We consider the Bayesian setting, where the $M$ alternatives $\tau_1, \ldots, \tau_M$ occur with (classical) probabilities $p_1, \ldots, p_M$, respectively.

A $M$-ary hypothesis test is a POVM $\Pc\!\triangleq\!\{ \Pi_{1},  \Pi_{2}, \ldots, \Pi_{M}\!\}$, $\sum \Pi_i = \openone$.
The test decides the alternative $\tau_i$ when the measurement with respect to $\Pc$ has outcome $i$.
The probability that the test $\Pc$ decides $\tau_j$ when $\tau_i$ is the true underlying state is thus $\tr \bigl(\tau_i \Pi_{j} \bigr)$ and the average error probability is
\begin{align}
  \epsilon(\Pc) \triangleq  1 - \sum_{i=1}^M p_i \tr\left(\tau_i \Pi_i\right).
  \label{eqn:mht-eps}
\end{align}
We define the minimum average error probability as
\begin{align}
  \epsilon^{\star} \triangleq \min_{\Pc} \epsilon(\Pc).
  \label{eqn:mht-mineps}
\end{align}
The test $\Pc$ minimizing \refE{mht-mineps} has no simple form in general.

\begin{lemma}[Holevo-Yuen-Kennedy-Lax conditions]\label{lem:Pstar}A test $\Pc^{\star} = \{ \Pi_{1}^{\star},  \ldots, \Pi_{M}^{\star} \}$ minimizes~\refE{mht-mineps} if and only if, for each $m=1,\ldots,M$,
\begin{align}
  \bigl( \Lambda(\Pc^{\star}) - p_m \tau_m \bigr)\Pi_m^{\star} 
    \,=\, \Pi_m^{\star}\bigl( \Lambda(\Pc^{\star}) - p_m \tau_m \bigr) &\,=\, 0,
    \label{eqn:mht-pistar1}\\
  \Lambda(\Pc^{\star}) - p_m \tau_m &\,\geq\, 0,
    \label{eqn:mht-pistar2}
\end{align}
where 
\begin{align}\label{eqn:Upsilon-def}
       \Lambda(\Pc^{\star}) &\triangleq \sum_{i=1}^{M} p_i \tau_i \Pi_i^{\star} = \sum_{i=1}^{M} p_i \Pi_i^{\star} \tau_i
\end{align}
is required to be self-adjoint\footnote{The operator $\Lambda(\Pc)$ takes a role of the Lagrange multiplier associated to the constraint $\sum \Pi_{i} = \openone$, which, involving self-adjoint operators requires $\Lambda$ to be self-adjoint.}.
\end{lemma}
\begin{IEEEproof}
This result follows from \cite[Th. 4.1, Eq. (4.8)]{holevo1973decision} or \cite[Th. I]{ykl1975optimum} after simplifying the resulting optimality conditions. 
\end{IEEEproof}


We next show an alternative characterization of the minimum error probability $\epsilon^{\star}$ as a function of a binary hypothesis test with certain parameters.
Let $\diag(\rho_1, \ldots, \rho_M)$ denote the block-diagonal matrix with diagonal blocks $\rho_1, \ldots, \rho_M$. We define \begin{align}
\Tc &\triangleq \diag \bigl(p_1 \tau_1, \ldots, p_M \tau_M\bigr),
\label{eqn:Tc-def}\\
\Dc(\mu_0) &\triangleq \diag \bigl(\tfrac{1}{M} \mu_0, \ldots, \tfrac{1}{M}\mu_0\bigr),
\label{eqn:Dc-def}
\end{align}
where $\mu_0$ is an arbitrary density operator acting on $\Hc$. Note that both $\Tc$ and $\Dc(\mu_0)$ are density operators themselves, 
since they are self-adjoint, positive semidefinite and have unit trace.

\begin{theorem}\label{thm:main-result} 
The minimum error probability of a Bayesian \mbox{$M$-ary} test discriminating among
states $\{\tau_1,\ldots,\tau_M\}$ with prior probabilities $\{p_1, \ldots, p_M\}$ satisfies
\begin{align}
  \epsilon^{\star} = \max_{\mu_0} \alpha_{\frac{1}{M}} \bigl(\Tc \,\|\, \Dc(\mu_0)\bigr),
  \label{eqn:main-result}
\end{align}
where $\Tc$ and $\Dc(\cdot)$ are given in \refE{Tc-def} and \refE{Dc-def}, respectively, and where the optimization is carried out over (unit-trace non-negative) density operators $\mu_0$.
\end{theorem} 
\begin{IEEEproof}
For any $\Pc = \{ \Pi_{1},  \Pi_{2}, \ldots, \Pi_{M} \}$ let us define the binary test $T' \triangleq \diag \left(\Pi_1, \ldots, \Pi_M\right)$. The error probabilities $\eps_{1|0}$ and $\eps_{0|1}$ of the test $T'$
are given by
\begin{align}
 \eps_{1|0}(T') &=  1 - \sum_{i=1}^M p_i \tr\left(\tau_i \Pi_i\right) = \epsilon(\Pc),
                \label{eqn:pi10-Pitilde}\\
 \eps_{0|1}(T') &= \frac{1}{M} \sum_{i=1}^M \tr\left(\mu_0 \Pi_i\right)\\
                 &= \frac{1}{M} \tr\left(\mu_0 \left(\sum\nolimits_{i=1}^M \Pi_i\right) \right)\\
                 &= \frac{1}{M} \tr\left( \mu_0 \right) = \frac{1}{M}.\label{eqn:pi01-Pitilde}
\end{align}
The (possibly suboptimal) test $T'$ has thus $\eps_{1|0}(T') = \epsilon(\Pc)$ where $\epsilon(\Pc)$ is defined in \refE{mht-eps} and $\eps_{0|1}(T') = \frac{1}{M}$. Therefore, using \refE{bht-alpha} and maximizing the resulting expression over $\mu_0$, we obtain
\begin{align}
  \epsilon(\Pc) \geq  \max_{\mu_0} \alpha_{\frac{1}{M}} \bigl(\Tc\,\|\, \Dc(\mu_0)\bigr).
  \label{eqn:lower-bound}
\end{align}

It remains to show that, for  $\Pc = \Pc^{\star}$ defined in Lemma~\ref{lem:Pstar}, the lower bound \refE{lower-bound} holds with equality.
To this end, we next demonstrate that the optimality conditions for $T_{\text{NP}}$ in Lemma~\ref{lem:NPlemma} and for $\Pc^{\star} = \{ \Pi_{1}^{\star},  \ldots, \Pi_{M}^{\star} \}$ in Lemma~\ref{lem:Pstar} are equivalent for a specific choice of $\mu_0$.

Let $\Pc^{\star} = \{ \Pi_{1}^{\star},  \ldots, \Pi_{M}^{\star} \}$ satisfy \refE{mht-pistar1}-\refE{mht-pistar2} and define
\begin{align}\label{eqn:mu0star}
   \mu_0^{\star} \triangleq \frac{1}{c_0^{\star}} \sum_{i=1}^M p_i \tau_i \Pi_{i}^{\star} = \frac{1}{c_0^{\star}}  \Lambda(\Pc^{\star}),
\end{align}
where $c_0^{\star}$ is a normalizing constant such that $\mu_0^{\star}$ is unit trace. 

Lemma \ref{lem:NPlemma} shows that the test $T_{\text{NP}}$ achieving \refE{lower-bound} is associated to the non-negative eigenspace of the matrix $\Tc-t \Dc(\mu_0)$. Given the block-diagonal structure of the matrix $\Tc-t \Dc(\mu_0)$, it is enough to consider binary tests $T_{\text{NP}}$ with block-diagonal structure. Then, we write $T_{\text{NP}} = \diag \left(T_1^{\text{NP}}, \ldots, T_M^{\text{NP}}\right)$.

For the choice $\mu_0 = \mu_0^{\star}$, and $t = M c_0^{\star}$, the $m$-th block-diagonal term in $\Tc-t \Dc(\mu_0)$ is given by
\begin{align}
   p_m\tau_m - \tfrac{t}{M} \mu_0
     &= p_m\tau_m - \Lambda(\Pc^{\star}).
     \label{eqn:blockTlambdaD-1}
\end{align}

The $m$-th block of the Neyman-Pearson test $T_m^{\text{NP}}$ must lie in the non-negative eigenspace of the matrix \refE{blockTlambdaD-1}. However, since \refE{mht-pistar2} implies that \refE{blockTlambdaD-1} is negative semidefinite, each block $T_m^{\text{NP}}$ can only lie in the null eigenspace of~\refE{blockTlambdaD-1}, $m=1,\ldots,M$.

According to \refE{mht-pistar1}, the operator $\Pi_{m}^{\star}$ belongs to the null eigenspace of \refE{blockTlambdaD-1}, $m=1,\ldots,M$. As a result, the choice 
\begin{align}
T_{\text{NP}}  
  &= \diag \left(\Pi_1^{\star}, \ldots, \Pi_M^{\star}\right)
\end{align}
satisfies the optimality conditions in Lemma~\ref{lem:NPlemma}. Moreover, since $\eps_{1|0}(T_{\text{NP}}) = \epsilon\bigl(\Pc^{\star}\bigr) = \epsilon^{\star}$ and $\eps_{0|1}(T_{\text{NP}}) = \frac{1}{M}$,  Lemma~\ref{lem:NPlemma} implies that \refE{main-result} holds with equality for $\mu_0 = \mu_0^{\star}$. Given the bound in \refE{lower-bound}, other choices of $\mu_0$ cannot improve the result, and Theorem \ref{thm:main-result} thus follows.
\end{IEEEproof}

\textit{Example 1:} Consider a hypothesis testing problem between $M=4$ (non-equiprobable) alternatives given by
\begin{align}\label{eqn:qht_example}
  \tau_1=\begin{bmatrix} 1 &0\\ 0 & 0\end{bmatrix},\quad
  \tau_2=\frac{1}{2}\begin{bmatrix} 1 &1\\ 1 & 1\end{bmatrix},\quad
  \tau_3=\frac{1}{2}\begin{bmatrix} 1 &-1\\-1 & 1\end{bmatrix},\quad
  \tau_4=\frac{1}{2}\begin{bmatrix} 1 &0\\ 0 & 1\end{bmatrix},
\end{align}
with prior probabilities $p_1=2/5$ and $p_2=p_3=p_4=1/5$. 
By solving \refE{mht-mineps}, we obtain $\epsilon^{\star} = 7/15 = 0.4\overline{6}$
which is attained by the measurement
$\Pc^{\star} = \{ \Pi_{1}^{\star},  \ldots, \Pi_{4}^{\star} \}$ with $\Pi_4^{\star}=0$ and
\begin{align}\label{eqn:qht_example-Pistar}
  \Pi_1^{\star}=\begin{bmatrix} 8/9 &0\\ 0 & 0\end{bmatrix},\quad
  \Pi_2^{\star}=\begin{bmatrix} 1/18 &1/6\\ 1/6 & 1/2\end{bmatrix},\quad
  \Pi_3^{\star}=\begin{bmatrix} 1/18 &-1/6\\ -1/6 & 1/2\end{bmatrix}.
\end{align}
Note that even when the dimension of the Hilbert space is $2$, there are $3$ active
measurement operators. Since they are positive semidefinite and  $\sum_{i=1}^4 \Pi_i^{\star} = \openone$,
the POVM is well defined. The POVM $\Pc^{\star}$ satisfies the optimality conditions from
Lemma~\ref{lem:Pstar} and therefore $\epsilon^{\star} = 0.4\overline{6}$
is the lowest average error probability for this testing problem. 

According to \refE{mu0star}, the auxiliary state
\begin{align}\label{eqn:qht_example_mu0star}
   \mu_0^{\star} = \frac{1}{c_0^{\star}} \sum_{i=1}^4 p_i \tau_i \Pi_{i}^{\star} = \frac{1}{4}\begin{bmatrix} 3 &0\\ 0 & 1\end{bmatrix},
\end{align}
is optimal in Theorem \ref{thm:main-result}. Indeed, it follows that\footnote{This computation can be done, e.g., by using \refE{NPcorollary-1} from Lemma~\ref{lem:NPcorollary} or by solving a semidefinite program.}
\begin{align}
  \alpha_{\frac{1}{4}} \bigl(\Tc \,\|\, \Dc(\mu_0^{\star})\bigr) = 0.4\overline{6} = \epsilon^{\star}.
\end{align}
Other choices of $\mu_0$ yield a lower bound on the average error probability $\epsilon^{\star}$.
For example, considering $\mu_0$ the average state for this testing problem,
\begin{align}\label{eqn:qht_example_mu0}
  \mu_0 = \sum_{m=1}^4 p_m \tau_m = \begin{bmatrix} 0.7 &0\\ 0 & 0.3\end{bmatrix},
\end{align}
yields
\begin{align}\label{eqn:qht_alpha_example_mu0}
  \alpha_{\frac{1}{4}} \bigl(\Tc \,\|\, \Dc(\mu_0)\bigr) \approx 0.4571 < 0.4\overline{6} = \epsilon^{\star}.
\end{align}

Theorem~\ref{thm:main-result} thus provides an alternative expression for the error probability $\epsilon^{\star}$ for the optimal choice of the auxiliary state, and a lower bound for other choices of $\mu_0$.
Combining Theorem~\ref{thm:main-result} and Lemma~\ref{lem:NPcorollary}, we obtain an alternative characterization for $\epsilon^{\star}$ based on information-spectrum measures.

\begin{corollary}\label{cor:tight-spectrum} 
The minimum error probability of an \mbox{$M$-ary} test discriminating among states $\{\tau_1,\ldots,\tau_M\}$ with prior classical probabilities $\{p_1, \ldots, p_M\}$ satisfies
\begin{align}
  \epsilon^{\star} = \max_{\mu_0, t\geq 0} \left\{ \sum_{i=1}^{M} p_i \tr\Bigl( \tau_i \bigl\{ p_i \tau_i - t \mu_0 \leq 0 \bigr\} \Bigr)- t \right\}\!.
\label{eqn:tight-spectrum}
\end{align}
where the optimization is carried out over (unit-trace non-negative) density operators $\mu_0$ acting on $\Hc$, and over the scalar threshold $t \geq 0$.
\end{corollary} 
\begin{IEEEproof}
Applying the lower bound \refE{NPcorollary-2} from Lemma~\ref{lem:NPcorollary} to the identity \refE{main-result}, and using  the definitions of $\Tc$ in \refE{Tc-def} and $\Dc(\cdot)$ in \refE{Dc-def}, it yields, for any $\mu_0$, $t'\geq 0$,
\begin{align}
  \epsilon^{\star} \geq \sum_{i=1}^{M} p_i \tr\Bigl( \tau_i \bigl\{ p_i \tau_i - \tfrac{t'}{M} \mu_0 \leq 0 \bigr\} \Bigr)- \tfrac{t'}{M}.
\label{eqn:lemma-hn-2}
\end{align}

It remains to show that there exist $\mu_0$ and $t'\geq 0$ such that \refE{lemma-hn-2} holds with equality.
In particular, let us choose $\mu_0 = \mu_0^{\star}$ defined in~\refE{mu0star}, and $t' = M c_0^{\star}$ where
   $c_0^{\star} = \sum_{i=1}^M p_i \tr(\tau_i \Pi_{i}^{\star})$
is the normalizing constant from~\refE{mu0star}.

For this choice of $\mu_0$ and $t'$, the projector spanning the negative semidefinite eigenspace of the operator \mbox{$p_i\tau_i\!-\!\frac{t'}{M} \mu_0$} can be rewritten as
\begin{align}
 \Bigl\{ p_i \tau_i -\tfrac{t'}{M} \mu_0 \leq 0 \Bigr\}
&= \bigl\{ p_i \tau_i - \Lambda(\Pc^{\star}) \leq 0 \bigr\}
= \openone, \label{eqn:trivialproj}
\end{align}
where the last identity follows from \refE{mht-pistar2}.
The right-hand side of \refE{lemma-hn-2} thus becomes
\begin{align}
\sum_{i=1}^{M} p_i \tr( \tau_i ) - \frac{t'}{M} 
\,=\, 1 - \frac{t'}{M}.
\label{eqn:lemma-hn-3}
\end{align}
The result follows since $\tfrac{t'}{M}\!=\! c_0^{\star}\!=\!\sum_{i} p_i \tr(\tau_i\Pi_{i}^{\star})\!=\!1\!-\!\epsilon^{\star}$.
\end{IEEEproof}

\begin{figure}[t]%
\centering\input{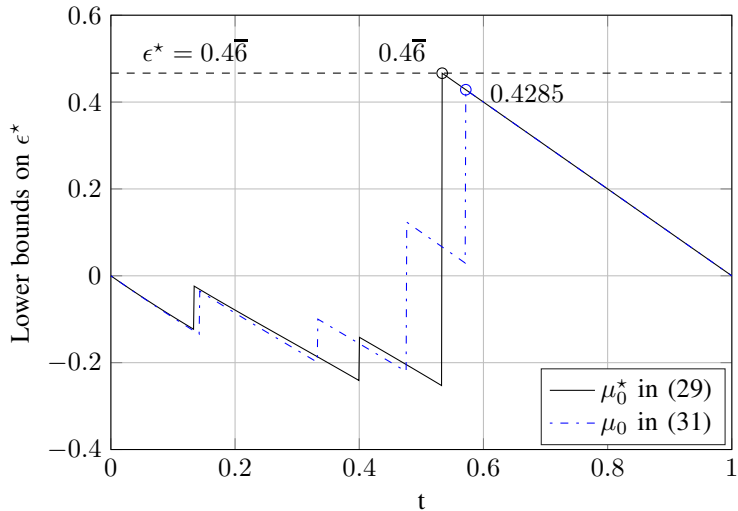}
\vspace{-2mm}\caption{Minimum error probability $\epsilon^{\star}$ (horizontal dashed line) for the hypothesis testing problem described in \refE{qht_example}, compared with the lower bound that follows from \refE{tight-spectrum} for fixed values of $t$ and $\mu_0$.}
\label{fig:example-quantumVH-tight}
\end{figure}

For illustration, let us consider again the testing problem from Example 1, c.f. \refE{qht_example}.
Figure~\ref{fig:example-quantumVH-tight} shows the objective of \refE{tight-spectrum}
as a function of $t$ for the auxiliary state $\mu_0=\mu_0^{\star}$ in \refE{qht_example_mu0star} and for the value of 
$\mu_0$ given in \refE{qht_example_mu0}. We can see that considering $\mu_0=\mu_0^{\star}$,
after maximization over $t$, yields the exact error probability $\epsilon^{\star}= 0.4\overline{6}$.
In contrast, considering the value of $\mu_0$ in \refE{qht_example_mu0}, it yields a strict
lower bound with a largest value of $0.4285$, approximately.
Comparing this value with \refE{qht_alpha_example_mu0}, we conclude that by fixing a suboptimal auxiliary state $\mu_0$, the right-hand side of \refE{main-result} from Theorem \ref{thm:main-result} yields tighter bounds than \refE{tight-spectrum} from Corollary \ref{cor:tight-spectrum}. This could be expected as the expression in Corollary \ref{cor:tight-spectrum} follows from a weakening of \refE{main-result}.

We recall from the proofs of both Theorem \ref{thm:main-result} and Corollary \ref{cor:tight-spectrum} that a density operator $\mu_0$ maximizing \refE{main-result} and \refE{tight-spectrum} is 
\begin{align}
   \mu_0^{\star} = \frac{1}{c_0^{\star}} \sum_{i=1}^M p_i \tau_i \Pi_{i}^{\star},
\end{align}
for some $\Pc^{\star} = \{ \Pi_{1}^{\star},  \ldots, \Pi_{M}^{\star} \}$ satisfying the conditions in Lemma \ref{lem:Pstar} and where $c_0^{\star}$ is a normalizing constant. Hence, the optimal $M$-ary hypothesis test $\Pc^{\star}$ characterizes the optimal $\mu_0$. Conversely, the optimal $\mu_0$ is precisely the Lagrange multiplier associated to the minimization in \refE{mht-mineps}, after an appropriate re-scaling.

While the expressions in Theorem \ref{thm:main-result} and Corollary \ref{cor:tight-spectrum} are not easier to
compute than the exact error probability, we show in the next section that they can be used to determine the tightness of several converse bounds in the context of reliable communication over classical-quantum channels.


\section{Classical-Quantum Channels}\label{sec:cq-channels}

We consider the channel coding problem of transmitting  $M$ equiprobable messages\footnote{While the results from Section~\ref{sec:mary} were derived for discrimination among non-equiprobable alternatives, in the remainder of this paper we consider the channel coding problem with equiprobable messages for clarity of exposition.} over a one-shot classical-quantum channel $x \to W_x$, with $x\in\Xc$ and $W_x\in\Hc$.
A channel code is defined as a mapping from the message set $\{1,\ldots,M\}$ into a set of $M$ codewords $\Cc = \{x_1,\ldots,x_M\}$.  For a source message $m$, the decoder receives the associated density operator $W_{x_m}$ and must decide on the transmitted message.

With some abuse of notation, for a fix code, sometimes we shall write $W_m \triangleq W_{x_m}$. The minimum error probability for a code $\Cc$ is then given by
\begin{align}
  \Pe(\Cc) &\triangleq
  \min_{\{\Pi_1,\ldots,\Pi_M\}} \left\{ 1 - \frac{1}{M} \sum_{m=1}^M \tr\bigl(W_m \Pi_m\bigr) \right \}.
  \label{eqn:PeC}
\end{align}

This problem corresponds precisely to the $M$-ary quantum hypothesis testing problem described in
Section \ref{sec:mary}.
In contrast to the classical setting, in which \refE{PeC} is minimized by the maximum likelihood
decoder, the minimizer of \refE{PeC} corresponds to any POVM satisfying the optimality conditions
from Lemma~\ref{lem:Pstar}.

A direct application of Theorem \ref{thm:main-result} yields an alternative expression for $\Pe(\Cc)$.
Let $P$ denote a (classical) distribution over the input alphabet $\Xc$ and define
\begin{align}
  PW \triangleq \sum\nolimits_{x\in\Xc} P(x)\Bigl( \left| x \rangle \langle x \right| \otimes W_x \Bigr),
  \label{eqn:PW_def}\\
  P \otimes \mu \triangleq \Bigl(\sum\nolimits_{x\in\Xc} P(x) \left| x \rangle \langle x \right|\Bigr) \otimes \mu.
  \label{eqn:Pmu_def}
\end{align}
We denote by $P_{\Cc}$, the input distribution induced by the codebook $\Cc$, hence
$P_{\Cc}W = \frac{1}{M} \sum_{x\in\Cc} \bigl( \left| x \rangle \langle x \right| \otimes W_x \bigr)$
and $P_{\Cc} \otimes \mu = \bigl(\frac{1}{M}\sum_{x\in\Cc} \left| x \rangle \langle x \right|\bigr) \otimes \mu$.
Using the alternative expression introduced in Theorem \ref{thm:main-result} we obtain the following result.
\begin{theorem}[Meta-converse]\label{thm:meta-converse}
Let $\Cc$ be any codebook of cardinality $M$ for a channel $W_x\in\Hc$.  Then,
\begin{align}
  \Pe(\Cc)
  &=  \sup_{\mu} \Bigl\{ \alpha_{\frac{1}{M}} \bigl(P_{\Cc}W \,\|\, P_{\Cc} \otimes \mu\bigr) \Bigr\}
  \label{eqn:alpha-cc}\\
  &\geq \inf_{P} \sup_{\mu} \Bigl\{ \alpha_{\frac{1}{M}} \bigl(PW \,\|\, P \otimes \mu\bigr) \Bigr\}.
  \label{eqn:meta-converse}
\end{align}
where the maximization is over auxiliary states~$\mu\in\Hc$, and the minimization is over (classical) input distributions~$P$.
\end{theorem}
\begin{IEEEproof}
The identity \refE{alpha-cc} is a direct application of \refE{main-result} in Theorem \ref{thm:main-result}.
The relaxation \refE{meta-converse} follows by minimizing \refE{alpha-cc} over all input distributions, not necessarily induced by a codebook.
\end{IEEEproof}

The right-hand-side of~\refE{alpha-cc} coincides with the finite block-length converse bound by Matthews and Wehner \cite[Eq.~(45)]{matthews2014finite}, particularized for a classical-quantum channel with an input state induced by the codebook $\Cc$. The lower bound \refE{meta-converse} corresponds to \cite[Eq.~(46)]{matthews2014finite} specialized to the  classical-quantum setting (see also \cite[Sec. 4.6]{Hayashi06} for a direct derivation for classical quantum channels).
The classical analogous of \refE{meta-converse} is usually referred to as meta-converse bound, since several converse bounds in the literature can be derived from it. As it is the case in the classical-quantum setting, in the following we shall refer to this result as \textit{meta-converse}.

Theorem~\ref{thm:meta-converse} implies that the quantum generalization of the meta-converse bound proposed by Matthews and Wehner in \cite[Eq.~(45)]{matthews2014finite} is tight for a fixed codebook $\Cc$. By fixing $\mu$ to be the state induced at the system output, the lower bound \refE{meta-converse} recovers the converse bound by Wang and Renner \cite[Th.~1]{wangrenner2012}. This bound is not tight in general since (i) the minimizing $\px$ does not need to coincide with the input state induced by the best codebook, and (ii) the choice of $\mu_0$ in \cite[Th. 1]{wangrenner2012} does not maximize the resulting bound in general.

Using the characterization from Corollary~\ref{cor:tight-spectrum}, the error probability $\Pe(\Cc)$ can be equivalently written as
\begin{align}
  \Pe(\Cc)\!=\! \max_{\mu_0, t\geq 0} \left\{ \frac{1}{M} \sum_{x\in\Cc}\tr\Bigl( W_x \bigl\{ W_x\!-\!t\mu_0 \leq 0 \bigr\} \Bigr)- \frac{t}{M} \right\}\!.
\label{eqn:tight-hn}
\end{align}
The objective of the maximization in~\refE{tight-hn} coincides with the information-spectrum bound \cite[Lemma~4]{hayashi2003general}. Then, \refE{tight-hn} shows that the Hayashi-Nagaoka lemma yields the exact error probability for a fixed code, after optimizantion over the free parameters $\mu_0$, $t\geq 0$.


\section{Quasi-Perfect Codes}\label{sec:qp-codes}

While the alternative expressions \refE{alpha-cc} and \refE{tight-hn} yield the
exact error probability, they still depend on the codebook~$\Cc$.
To obtain a practical converse bound, these expressions need to be minimized over a family of codes or input distributions. One practical converse bound is the relaxation in~\refE{meta-converse} which yields a practical lower bound that can be evaluated in several cases of interest. 
Since the meta-converse bound \refE{meta-converse} is a relaxation, it does not coincide with the exact error probability in general. Nevertheless, we next show that this is still the case for a the family of codes defined in this section. 

We consider the classical-quantum channel model introduced in \refS{cq-channels}.
For the classical-quantum channel $\{W_x\}$, $x\in\Xc$, and parameters $t\in\RR$ and $\mu\in\Hc$, we define
\begin{align}
  \Ec_{x}(t,\mu) &\triangleq \bigl\{ W_{x} - t \mu \geq 0 \bigr\},  
  \label{eqn:Ex_def}\\
  F_{x}(t,\mu) &\triangleq \tr\bigl( W_x \Ec_{x}(t,\mu) \bigr),
  \label{eqn:Fx_def}\\
  G_{x}(t,\mu) &\triangleq \tr\bigl( \mu\,\Ec_{x}(t,\mu) \bigr),
  \label{eqn:Gx_def}
\end{align}
and we consider the following family of symmetric channels.
\begin{definition} \label{def:symmetric}
 A channel $\{W_x\}$, $x\in\Xc$, is \textit{symmetric} with respect to $\mu\in\Hc$
 if $F_{x}(t,\mu)$ does not depend on $x\in\Xc$ for any $t \in \RR$, i.e,
\begin{align}\label{eqn:symmetric}
  F_{x}(t,\mu)  = F(t,\mu),\quad  \forall x\in\Xc,\ t \in \RR. 
\end{align}
\end{definition}
Using \refE{symmetric}, it can be shown that $G_{x}(t,\mu) = G(t,\mu)$ does not depend on $x$
for any channel which is symmetric with respect to $\mu$.
Similarly to \refE{Ex_def}-\refE{Gx_def}, we define
\begin{align}
  \Ec^{\bullet}_{x}(t,\mu) &\triangleq \bigl\{ W_{x} - t \mu > 0 \bigr\},  
  \label{eqn:Eix_def}\\
  F^{\bullet}_{x}(t,\mu) &\triangleq \tr\bigl( W_x \Ec_{x}^{\bullet}(t,\mu) \bigr),
  \label{eqn:Fix_def}\\
  G^{\bullet}_{x}(t,\mu) &\triangleq \tr\bigl( \mu\,\Ec_{x}^{\bullet}(t,\mu) \bigr),\label{eqn:Gix_def}
\end{align}
and, for a symmetric channel, $F_{\bullet}(\cdot) \triangleq F^{\bullet}_{x}(\cdot)$, $G_{\bullet}(\cdot) \triangleq G^{\bullet}_{x}(\cdot)$.
\begin{definition} \label{def:quasiperfect}
A code $\Cc$ is \textit{perfect} for a classical-quantum channel $\{W_x\}$, if there exists a scalar $t$ and a state $\mu\in\Hc$ such that 
the projectors $\bigl\{\Ec_{x}(t,\mu) \bigr\}_{x \in \Cc}$ are orthogonal to each other and $\sum_{x\in\Cc} \Ec_{x}(t,\mu) = \openone$. More generally, a code is \textit{quasi-perfect} if  there exists $t$ and $\mu\in\Hc$ such that 
the projectors $\bigl\{  \Ec_{x}^\bullet(t,\mu) \bigr\}_{x \in \Cc}$ are orthogonal to each other and $\sum_{x\in\Cc} \Ec_{x}(t,\mu) \geq \openone$. 
\end{definition}

\textit{Example 2:} Let us consider the pure-state channel $x \rightarrow W_x = \ket{\varphi_x}\bra{\varphi_x} \in \Hc$, where the output space has $n$ dimensions. This channel is symmetric with respect to the maximally mixed state $\mu=\frac{1}{n}\openone$. To see this, note that
\begin{align}
  F_{x}(t,\mu)
  &= \tr\bigl( W_x \bigl\{ W_{x} - \tfrac{t}{n}\openone \geq 0 \bigr\} \bigr)\\
  &= \bra{\varphi_x} \bigl\{ \ket{\varphi_x}\bra{\varphi_x} - \tfrac{t}{n}\openone \geq 0 \bigr\} \ket{\varphi_x}.
\end{align}
The projector $\bigl\{ \ket{\varphi_x}\bra{\varphi_x} - \tfrac{t}{n}\openone \geq 0 \bigr\} = \openone$ for $t<0$; the only non-negative eigenvalue of \mbox{$\bigl\{ \ket{\varphi_x}\bra{\varphi_x} - \tfrac{t}{n}\openone \geq 0 \bigr\}$} is associated to the eigenvector $\ket{\varphi_x}$ for $0\leq t\leq n$; and all the eigenvalues are negative for $t > n$, hence  $\bigl\{ \ket{\varphi_x}\bra{\varphi_x} - \tfrac{t}{n}\openone \geq 0 \bigr\} = 0$. Then, we conclude that 
\begin{align}
  F_{x}(t,\mu) = \begin{cases} 1, & t\leq n,\\
                               0, & t > n,
  \end{cases}
\end{align}
which does not depend on $x$.

Then, according to Definition~\ref{def:quasiperfect}, a code $\Cc$ with $M=n$ orthogonal pure states is perfect for this channel with parameters $t=n$ and $\mu=\frac{1}{n}\openone$, since the projectors $\Ec_{x}\bigl(n,\frac{1}{n}\openone\bigr) = \bigl\{ \ket{\varphi_x}\bra{\varphi_x} - \openone \geq 0 \bigr\} = \ket{\varphi_x}\bra{\varphi_x}$ are orthogonal for $x\in\Cc$, and they form a basis for $\Hc$.
Similarly, a code with $M \geq n$ is quasi-perfect for this channel with parameters $t=n$ and $\mu=\frac{1}{n}\openone$ provided that $\sum_{x\in\Cc} \ket{\varphi_x}\bra{\varphi_x} \geq \openone$, since the interiors $\Ec_{x}^\bullet\bigl(n,\frac{1}{n}\openone\bigr)= \bigl\{ \ket{\varphi_x}\bra{\varphi_x} - \openone > 0 \bigr\} = 0$, hence they are orthogonal. For $M < n$, the codes for this channel and the auxiliary state $\mu=\frac{1}{n}\openone$ are neither perfect nor quasi-perfect.


To avoid ambiguities, we shall denote by $\bar{t}$ the smallest value of $t$ such that the projectors $\bigl\{  \Ec_{x}^\bullet(t,\mu) \bigr\}_{x \in \Cc}$ are orthogonal to each other for a certain code $\Cc$. We shall refer to $\bar{t}$ as the \textit{packing radius} of the code $\Cc$ with respect to state $\mu$.
Similarly, we define $\epsilon \geq 0$ as the smallest value such that $\sum_{x\in\Cc} \Ec^{\epsilon}_{x}(\bar{t},\mu) \geq \openone$, where 
\begin{align}
  \Ec^{\epsilon}_{x}(\bar{t},\mu) \triangleq \bigl\{ W_{x} - \bar{t}\mu \geq - \epsilon \openone \bigr\}.
\end{align}
We denote $\epsilon$ as the \textit{optimality gap} of the code for a reason that will became apparent in the sequel. Note that for perfect and quasi-perfect codes the packing radius $\bar{t}$ is the value of $t$ appearing in Definition~\ref{def:quasiperfect} and the optimality gap is $\epsilon=0$.

The next result provides an alternative expression for the error probability
of perfect and quasi-perfect codes.

\begin{theorem}[Error probability of quasi-perfect codes]
\label{thm:Pe-quasiperfect}
Let the channel $\{W_x\}$ be symmetric with respect to $\mu$ and let $\Cc$ be perfect or quasi-perfect with parameters $t$ and $\mu$. Then,
\begin{align}
  \Pe(\Cc) &= 1-F_{\bullet}(t,\mu) + t\bigl(G_{\bullet}(t,\mu)-|\Cc|^{-1}\bigr),\label{eqn:Pe-quasiperfect}
\end{align}
where $|\Cc|$ denotes the cardinality of the codebook $\Cc$. Conversely, the right-hand side of equation \refE{Pe-quasiperfect} is an strict lower bound to the error probability if the code $\Cc$ is not quasi-perfect with parameters $t$ and $\mu$.
\end{theorem}
\begin{IEEEproof}
Let $\Cc = \{x_1,\ldots,x_M\}$ be an arbitrary code for the (symmetric) channel $\{W_x\}$. Let $\bar{t}$ be the packing radius of $\Cc$ with respect to the auxiliary state $\mu$, and let $\epsilon$ be the corresponding optimality gap.

We define an orthogonal basis $\{E_i\}$ such that
\begin{align}
    \Ec^{\bullet}_{x}(\bar{t},\mu) = \sum_{i \in {\Ic}(x)} E_i , \quad\text{for all } x\in\Cc. \label{eqn:Ix-def}
\end{align}
Here, $\Ic(x)$ denotes the set of basis indexes ``closest'' to the codeword $x\in\Cc$. This decomposition is guaranteed to exist since $\bar{t}$ being the packing radius of code code $\Cc$ and state $\mu$ implies that the projectors $\bigl\{  \Ec_{x}^\bullet(\bar{t},\mu) \bigr\}_{x \in \Cc}$ are orthogonal to each other.
We also define the set of indexes that do not belong to any of the projectors \refE{Ix-def} as
\begin{align}
\Ic_0 \triangleq \Bigl\{ i\,|\, i\notin \bigcup\nolimits_{x\in\Cc}\Ic(x)\Bigr\}.\label{eqn:I0-def}
\end{align}

As we did in the sets $\Ic(x)$, we shall assign the basis indexes in $\Ic_0$ to the different codewords. To this end, for each basis element $E_i$ and codeword $x\in\Xc$, we define the metric $\epsilon_i(x)$ such that
\begin{align}
   W_{x} E_i = \left(\bar{t}\mu - \epsilon_i(x)\openone\right) E_i.
   \label{eqn:epsi-def}
\end{align}
Intuitively, small values of $\epsilon_i(x)$ indicate that the basis element $E_i$ is ``close'' to the codeword $x\in\Xc$. 
We now assign the basis indexes in $\Ic_0$ to the codewords based on this metric. With some abuse of notation we define
\begin{align}
 \Ic_{0}(x) \triangleq \Bigl\{ i\in \Ic_0\;\Big|\; x = \arg \min_{x\in\Cc} \epsilon_i(x) \Bigr\}\label{eqn:Ieps-defd}
\end{align}
In case there is more than one such $x = \arg \min_{x\in\Cc} \epsilon_i(x)$ we assign one of them arbitrarily. Then, $\Ic(x)$, $x\in\Xc$, and $\Ic_{0}(x)$, $x\in\Xc$, define a partition of the basis indexes.

We consider the decoder $\Pc = \{ \Pi_{1},  \ldots, \Pi_{M} \}$ with projectors
\begin{align}
  \Pi_{m} = \Ec^{\bullet}_{x_m}(\bar{t},\mu) + &\Ec^{\circ}_{x_m}(\bar{t},\mu),\quad m=1,\ldots,M, \label{eqn:Pim-def}
\end{align}
where $\Ec^{\bullet}_{x}(\bar{t},\mu)$ is defined in \refE{Ix-def} and
\begin{align}
\Ec^{\circ}_{x}(\bar{t},\mu) \triangleq \sum_{i \in \Ic_0(x)}  E_i , \quad\text{for all } x\in\Cc.\label{eqn:Ieps-def}
\end{align}

According to the definitions of packing radius and optimality gap, it follows that $\Ec^{\bullet}_{x_m}(\bar{t},\mu) \leq \Pi_{m} \leq \Ec^{\epsilon}_{x}(\bar{t},\mu)$.  If we define $\epsilon_i \triangleq \min_{x\in\Cc} \epsilon_i(x)$, the condition $\sum_{x\in\Cc} \Ec^{\epsilon}_{x}(\bar{t},\mu) \geq \openone$ implies that the optimality gap is given by $\epsilon = \max_i \epsilon_i$. Substituting \refE{Ix-def} and \refE{Ieps-def} in \refE{Pim-def}, we obtain 
\begin{align}
  \Pi_{m} = \sum_{i \in \Ic(x_m)}  E_i+\sum_{i \in \Ic_0(x_m)} E_i.\label{eqn:Pi_Ei}
\end{align}
Since $\Ic(x)$ and $\Ic_{0}(x)$ define a partition of the basis indexes,
it follows that 
\begin{align}
 \sum_{m=1}^M \Pi_{m}=\openone,\label{eqn:Pim-sum}
 \end{align}
as required.

We next show that this decoder satisfies the Holevo-Yuen-Kennedy-Lax conditions from Lemma~\ref{lem:Pstar} and therefore it minimizes \refE{PeC}. The basis $\{E_i\}$ jointly diagonalizes the projectors $\bigl\{  \Ec_{x}^\bullet(\bar{t},\mu), \Ec_{x}^\circ(\bar{t},\mu) \bigr\}_{x \in \Cc}$. Indeed,
\begin{align}
\Lambda(\Pc) 
&= \frac{1}{M} \sum_{\ell=1}^{M}  W_\ell \Pi_\ell \\
&= \frac{1}{M} \sum_{\ell=1}^{M}  W_\ell \left(  \Ec^{\bullet}_{x_\ell}(\bar{t},\mu) + \Ec^{\circ}_{x_\ell}(\bar{t},\mu) \right)\\
&= \frac{1}{M} \sum_{\ell=1}^{M}  W_\ell \Ec^{\bullet}_{x_\ell}(\bar{t},\mu) +  \frac{1}{M} \sum_{\ell=1}^{M} \sum_{i \in \Ic_0(x_\ell)} \left(\bar{t}\mu -\epsilon_i\openone\right) E_i,\label{eqn:LambdaQP}\\
&= \frac{1}{M} \sum_{\ell=1}^{M} \sum_{i \in \Ic(x_\ell)} W_\ell E_i
  + \frac{1}{M}  \sum_{i \in \Ic_0}  \left(\bar{t}\mu -\epsilon_i\openone\right) E_i
  \label{eqn:LambdaQP2}
\end{align}
where \refE{LambdaQP} follows from \refE{epsi-def}, since $E_i$ belonging to the subspace $\Ec^{\circ}_{x_\ell}(\bar{t},\mu)$ implies that  $W_\ell E_i = \left(\bar{t}\mu -\epsilon_i\openone\right) E_i$ when $i \in \Ic_{0}(x_\ell)$. Using \refE{LambdaQP2}, it follows that
\begin{align}
\left(\Lambda(\Pc) - \frac{1}{M} W_m  \right) \Pi_m
&= \frac{1}{M} \sum_{\ell=1}^{M} \sum_{i \in \Ic(x_\ell)} W_\ell E_i\Pi_m
  + \frac{1}{M}  \sum_{i \in \Ic_0}  \left(\bar{t}\mu -\epsilon_i\openone\right) E_i\Pi_m  - \frac{1}{M} W_m \Pi_m
  \label{eqn:LambdaQPminusWmPim-0}\\
&= \frac{1}{M} \sum_{\ell\neq m} \sum_{i \in \Ic(x_\ell)} W_\ell E_i\Pi_m
  + \frac{1}{M}  \sum_{i \in \Ic_0\setminus \Ic_0(x_m) }  \left(\bar{t}\mu -\epsilon_i\openone\right) E_i\Pi_m.\label{eqn:LambdaQPminusWmPim}
\end{align}
where in the second step, we used \refE{Pi_Ei} in the last term of \refE{LambdaQPminusWmPim-0} and simplified the resulting expression.
Noting that, for $\ell\neq m$, it follows that $E_i \Pi_m = 0$ for any $i \in \Ic(x_\ell) \cup \Ic_0 \setminus \Ic_0(x_m)$, we conclude that 
$\left(\Lambda(\Pc) - \frac{1}{M} W_m  \right) \Pi_m = 0$.
Following analogous steps we show that
$\Pi_m \left(\Lambda(\Pc^{\star}) - \frac{1}{M} W_m  \right) =0$ and hence the decoder satisfies the optimality condition \refE{mht-pistar1}.

On the other hand, using \refE{LambdaQP2}, since $\sum_{i} E_i = \openone$, we obtain
\begin{align}
\Lambda(\Pc) - \frac{1}{M} W_m
&= \frac{1}{M} \sum_{\ell=1}^{M} \sum_{i \in \Ic(x_\ell)} W_\ell E_i
  + \frac{1}{M}  \sum_{i \in \Ic_0}  \left(\bar{t}\mu -\epsilon_i\openone\right) E_i-\frac{1}{M}\sum_{i}  W_m E_i\\
&=\frac{1}{M} \sum_{\ell\neq m} \sum_{i \in \Ic(x_\ell)} W_\ell E_i
  + \frac{1}{M}  \sum_{i \in \Ic_0\setminus \Ic_0(x_m) }  \left(\bar{t}\mu -\epsilon_i\openone\right) E_i\notag\\
&\quad-\frac{1}{M}\sum_{i \in \cup_{\ell\neq m}\Ic(x_\ell)} W_mE_i-\frac{1}{M}\sum_{i \in \Ic_0\setminus \Ic_0(x_m) } W_m E_i
  \label{eqn:LambdaQPminusWm}
\end{align}
For $i \in \Ic(x_\ell)$, using the definition of the projector $\Ec^{\bullet}_{x}(t,\mu)$ in \refE{Eix_def}, it follows that $W_\ell E_i > \bar{t}\mu E_i$. Similarly, for $i \in \cup_{\ell\neq m}\Ic(x_\ell)$, it follows that $W_m E_i \leq \bar{t}\mu E_i$. In addition, using \refE{Ieps-defd} and \refE{Ieps-def}, it follows that $W_m E_i \leq \left(\bar{t}\mu -\epsilon_i\openone\right) E_i$ for $i \in \Ic_0\setminus \Ic_0(x_m)$. Then, from \refE{LambdaQPminusWm}, we conclude that $\Lambda(\Pc^{\star}) - \frac{1}{M} W_m$ is lower bounded by
\begin{align}
\frac{1}{M} \sum_{\ell\neq m} \sum_{i \in \Ic(x_\ell)}\!\bar{t}\mu E_i
  + \frac{1}{M}  \sum_{i \in \Ic_0\setminus \Ic_0(x_m) }\!\left(\bar{t}\mu -\epsilon_i\openone\right) E_i -\frac{1}{M}\sum_{i \notin \cup_{\ell\neq m}\Ic(x_\ell)}\!\bar{t}\mu E_i-\frac{1}{M}\sum_{i \in \Ic_0\setminus \Ic_0(x_m)}\! \left(\bar{t}\mu -\epsilon_i\openone\right) E_i= 0,
\end{align}
and therefore \refE{mht-pistar2} holds. 

As the decoder $\Pc = \{ \Pi_{1},  \ldots, \Pi_{M} \}$ satisfies the optimality conditions from Lemma~\ref{lem:Pstar}, it minimizes \refE{PeC}. Then, combining  \refE{Upsilon-def} and \refE{PeC}, we obtain that the error probability of this code can be rewritten as
\begin{align}
  \Pe(\Cc) &=  1 - \tr\bigl( \Lambda(\Pc^{\star}) \bigr)\\
    &=  1 -
\frac{1}{M} \sum_{\ell=1}^{M}  W_\ell \Ec^{\bullet}_{x_\ell}(\bar{t},\mu)
  - \frac{1}{M} \sum_{i \in \Ic_0} \tr\bigl(\left(\bar{t}\mu -\epsilon_i\openone\right) E_i\bigr) \label{eqn:PeQP-0}\\
  &=  1 -
  \frac{1}{M} \sum_{m=1}^{M} F^{\bullet}_{x_m}(\bar{t},\mu)
  - \frac{\bar{t}}{M} \sum_{i \in \Ic_0} \tr\bigl( \mu E_i \bigr)
  + \frac{1}{M} \sum_{i \in \Ic_0} \epsilon_i,
  \label{eqn:PeQP-1}
\end{align}
where \refE{PeQP-0} follows from \refE{LambdaQP}, and in the last step we used \refE{Fix_def} and rearranged terms.

We now combine $\sum_{i} \tr\bigl( \mu E_i \bigr) = 1$, \refE{Gix_def} and \refE{I0-def} to obtain
\begin{align}
  \sum_{i \in \Ic_0} \tr\bigl( \mu E_i \bigr) + \sum_{m=1}^M G^{\bullet}_{x_m}(\bar{t},\mu) = 1.   \label{eqn:sumGxQP-1}
\end{align}
Multiplying both sides by $1/M$, noting that for a symmetric channel $G_{\bullet}(\bar{t},\mu) = G^{\bullet}_{x}(\bar{t},\mu)$ does not depend on $x$, from \refE{sumGxQP-1} it follows that
\begin{align}
  \frac{1}{M} \sum_{i \in \Ic_0} \tr\bigl( \mu E_i \bigr) = \frac{1}{M} - G_{\bullet}(\bar{t},\mu).   \label{eqn:sumGxQP-2}
\end{align}
Then, substituting \refE{sumGxQP-2} in \refE{PeQP-1}, and using $F_{\bullet}(\bar{t},\mu) = F^{\bullet}_{x}(\bar{t},\mu)$ and $M = |\Cc|$, we obtain
\begin{align}
  \Pe(\Cc) &= 1-F_{\bullet}(\bar{t},\mu) + \bar{t}\bigl(G_{\bullet}(\bar{t},\mu)-|\Cc|^{-1}\bigr)+ \frac{1}{M} \sum_{i \in \Ic_0} \epsilon_i.
  \label{eqn:Pe-gap-general-code}
\end{align}

Perfect and quasi-perfect codes have an optimality gap $\eps=\max_i\eps_i=0$. Therefore, if the code is quasi-perfect, then $\frac{1}{M} \sum_{i \in \Ic_0} \epsilon_i=0$ and $\Pe(\Cc)$ is given by the expression in the theorem. Conversely, if the code is not quasi-perfect, then at least one of the terms $\epsilon_i$ is greater than zero and the bound is strict.
\end{IEEEproof}

For the pure-state channel $x \rightarrow W_x = \ket{\varphi_x}\bra{\varphi_x}$ introduced in Example 2
above, let $t=n$ be the number of dimensions and $\mu=\frac{1}{n}\openone$ be the maximally mixed state.
Then, $F_{\bullet}(t,\mu)=G_{\bullet}(t,\mu)=0$ and using \refE{Pe-quasiperfect} we obtain that
for any perfect or quasi-perfect code $\Cc$ with cardinality $|\Cc|=M$, the error probability is given by
\begin{align}
  \Pe(\Cc) = 1 - \frac{n}{M}.\label{eqn:error-prob-pure-n}
\end{align}
Note that $\Pe(\Cc)$ is the average error probability of the code and that it does not describe
how the errors are distributed among the different messages.
It could happen that some of the projectors are inactive and the corresponding messages always
yield an error, and that some messages are decoded with no error.

We next show that perfect and quasi-perfect codes attain the meta-converse bound~\refE{meta-converse}
with equality. This result is based on the following auxiliary lemma.

\begin{lemma}\label{lem:alphabeta-decomposition}
Let $\rho_0 = PW$ and $\rho_1 = P \otimes \mu$ be defined in \refE{PW_def} and \refE{Pmu_def}, respectively.
Then, the optimal trade-off \refE{bht-alpha} for a hypothesis test between 
$\rho_0$ and $\rho_1$ satisfies
\begin{align}
\alpha_{\beta}\bigl(PW \,\|\, P \otimes \mu\bigr)
\,=\, \inf_{\substack{\{\beta_x'\}:\\ \beta = \sum_{x} P(x) \beta_x'}}
\sum_{x\in\Xc} P(x) \alpha_{\beta_x'} \bigl(W_x \,\|\, \mu\bigr).
\label{eqn:alphabeta-decomposition}
\end{align}
\end{lemma}
\begin{IEEEproof}
We consider Lemma~\ref{lem:NPcorollary} with $\rho_0\leftarrow PW$ and $\rho_1\leftarrow P \otimes \mu$. Then, using the block-diagonal structure of $PW$ and $P \otimes \mu$, the identity \refE{NPcorollary-1} yields
\begin{align}
\alpha_{\beta}\bigl(PW \,\|\, P \otimes \mu\bigr)
&= \sup_{t \geq 0} \Biggl\{ \sum_{x\in\Xc} P(x) \tr\bigl( W_x \bigl\{W_x - t \mu \leq 0\bigr\} \bigr)
 + t \Biggl( \sum_{x\in\Xc} P(x) \tr\bigl(\mu \bigl\{ W_x - t \mu  > 0\bigr\}\bigr) - \beta\Biggr) \Biggr\}
\label{eqn:alphabeta-decomposition-1}\\
&= \sup_{t \geq 0} \Biggl\{ \sum_{x\in\Xc} P(x) \biggl( \tr\bigl( W_x \bigl\{W_x - t \mu \leq 0\bigr\} \bigr)
 + t\Bigl(\tr\bigl(\mu \bigl\{ W_x - t \mu  > 0\bigr\}\bigr) - \beta_x' \Bigr) \biggr) \Biggr\}
\label{eqn:alphabeta-decomposition-2}
\end{align}
for any $\{\beta_x'\}$, $x\in\Xc$, such that $\sum_{x}P(x)\beta_x' = \beta$.

We relax the optimization \refE{alphabeta-decomposition-2} by 
letting the parameter $t$ be different for each $x$. Then, we obtain the following upper bound on $\alpha_{\beta}\bigl(PW \,\|\, P \otimes \mu\bigr)$,
\begin{align}
\alpha_{\beta}\bigl(PW \,\|\, P \otimes \mu\bigr)
&\leq  \sum_{x\in\Xc} P(x) \sup_{t_x \geq 0} \biggl\{ \tr\bigl( W_x \bigl\{W_x - t_x \mu \leq 0\bigr\} \bigr)
 + t_x\Bigl(\tr\bigl(\mu \bigl\{ W_x - t_x \mu > 0\bigr\}\bigr) - \beta_x' \Bigr) \biggr\}
\label{eqn:alphabeta-decomposition-3}\\
&= \sum_{x\in\Xc} P(x)  \alpha_{\beta_x'} \bigl(W_x \,\|\, \mu\bigr),\label{eqn:alphabeta-decomposition-4}
\end{align}
where in the last step we applied the identity \refE{NPcorollary-1} from Lemma~\ref{lem:NPcorollary} with $\rho_0\leftarrow W_x$ and $\rho_1\leftarrow \mu$. 
The bound \refE{alphabeta-decomposition-3}-\refE{alphabeta-decomposition-4} holds for any 
$\{\beta_x'\}$, $x\in\Xc$, such that $\sum_{x}P(x)\beta_x' = \beta$.
Then, to prove \refE{alphabeta-decomposition} it suffices to show that there exists
$\{\beta_x'\}$ satisfying $\sum_{x}P(x)\beta_x' = \beta$ and such that
\refE{alphabeta-decomposition-3} holds with equality. 

Indeed, the value of $t$ maximizing \refE{alphabeta-decomposition-2} induces the Neyman-Pearson test
\refE{NPlemma}, which due to the block-diagonal structure of the problem, can be decomposed into
the sub-tests
\begin{align}
  T_{x}' = \bigl\{ W_x - t \mu > 0\bigr\} + \theta^{0}_{x}.
\end{align}
Each of these subtests induces a type-I error probability $\alpha_x'$ and type-II error probability $\beta_x'$,
which, according to the NP lemma, satisfy  $\sum_{x}P(x)\alpha_x' = \alpha_{\beta}\bigl(PW \,\|\, P \otimes \mu\bigr)$
and $\sum_{x}P(x)\beta_x' = \beta$. 
It follows that, for this choice of $\{\beta_x'\}$, the optimization in \refE{alphabeta-decomposition-3} yields $t_x=t$ (as the $t$ parameter in the NP subtests is unique), and therefore \refE{alphabeta-decomposition-3} holds with equality. The result thus follows.
\end{IEEEproof}

Lemma~\ref{lem:alphabeta-decomposition} asserts that, for a binary hypothesis test between classical-quantum distributions, it is possible to express the optimal type-I error probability as a convex combination of that of disjoint sub-tests provided that the type-II error is optimally distributed among them.
The next result follows from combining Theorem~\ref{thm:Pe-quasiperfect} and Lemmas~\ref{lem:NPcorollary} and~\ref{lem:alphabeta-decomposition}.

\begin{theorem}[Quasi-perfect codes attain the meta-converse]
\label{thm:alpha-quasiperfect}
Let the channel $\{W_x\}$ be symmetric with respect to $\mu$ and let $\Cc$ be perfect or quasi-perfect with parameters $t$ and $\mu$. Then, for $M=|\Cc|$, 
\begin{align}
  \Pe(\Cc) &= \inf_{P} \sup_{\mu'} \alpha_{\frac{1}{M}} \bigl(PW \,\|\, P \otimes \mu'\bigr)
  \label{eqn:alpha-quasiperfect-1}\\
  &= \alpha_{\frac{1}{M}} \bigl(W_x \,\|\, \mu\bigr).\label{eqn:alpha-quasiperfect-2}
\end{align}
\end{theorem}
\begin{IEEEproof}
According to \refE{meta-converse} in Theorem \ref{thm:meta-converse}, the right-hand side of \refE{alpha-quasiperfect-1} is a lower bound to the error probability of any code. Then, to prove \refE{alpha-quasiperfect-1}, it suffices to show that the error probability of $\Cc$ coincides with this lower bound.
Using Lemma~\ref{lem:alphabeta-decomposition}, fixing the auxiliary state $\mu$ to that from Definition~\ref{def:quasiperfect}, we obtain
\begin{align}
 \inf_{P} \sup_{\mu'} \alpha_{\frac{1}{M}} \bigl(PW \,\|\, P \otimes \mu'\bigr) \geq \inf_{\substack{\{P(x),\beta_x\}:\\ \sum_{x} P(x) \beta_x = \frac{1}{M}}}
\sum_{x\in\Xc} P(x) \alpha_{\beta_x} \bigl(W_x \,\|\, \mu\bigr).
\label{eqn:alpha-quasiperfect-3}
\end{align}
Now, using \refE{NPcorollary-1} from Lemma~\ref{lem:NPcorollary}, letting $t'=t$, and using the definitions of $F^{\bullet}_{x}(t,\mu)$ and $G^{\bullet}_{x}(t,\mu)$, it follows that
\begin{align}
\alpha_{\beta_x} \bigl(W_x \,\|\, \mu\bigr)
&\geq 1- F^{\bullet}_{x}(t,\mu)  + t \bigl( G^{\bullet}_{x}(t,\mu) - \beta_x \bigr)\\
&= 1- F_{\bullet}(t,\mu)  + t \bigl( G_{\bullet}(t,\mu) - \beta_x \bigr),
\label{eqn:alpha-quasiperfect-4}
\end{align}
where in the last step we used that for symmetric channels, $F_{\bullet}(t,\mu) = F^{\bullet}_{x}(t,\mu)$ and $G_{\bullet}(t,\mu) = G^{\bullet}_{x}(t,\mu)$.

Then, using \refE{alpha-quasiperfect-4} in \refE{alpha-quasiperfect-3}, we obtain
\begin{align}
 \inf_{P} \sup_{\mu'} \alpha_{\frac{1}{M}} \bigl(PW \,\|\, P \otimes \mu'\bigr) &\geq\!
 \inf_{\substack{\{P(x),\beta_x\}:\\ \sum_{x} P(x) \beta_x = \frac{1}{M}}}
\hspace{-0.2mm}\Bigl(1- F_{\bullet}(t,\mu)  
+ t \Bigl( G_{\bullet}(t,\mu) -\!\textstyle\sum_{x}\!P(x) \beta_x \Bigr)\Bigr)\\
&= 1- F_{\bullet}(t,\mu)  
+ t \biggl( G_{\bullet}(t,\mu) - \frac{1}{M} \biggr)
\label{eqn:alpha-quasiperfect-5}
\end{align}
where in the second step we used the constraint $\sum_{x} P(x) \beta_x = \frac{1}{M}$ since the resulting objective does not depend on the optimization variables.

The right-hand side of \refE{alpha-quasiperfect-5} coincides with the
error probability of the quasi-perfect codes given in \refE{Pe-quasiperfect}. Then, using this observation and \refE{meta-converse}  we conclude that, whenever $\Cc$ is perfect or quasi-perfect, 
\begin{align}
   \Pe(\Cc) \leq 
 \inf_{P} \sup_{\mu'} \alpha_{\frac{1}{M}} \bigl(PW \,\|\, P \otimes \mu'\bigr) \leq  \Pe(\Cc),
\end{align}
and the meta-converse bound \refE{alpha-quasiperfect-1} must hold with equality.
Since \refE{Pe-quasiperfect} coincides with the lower bound \refE{alpha-quasiperfect-4} when $\beta_x=\frac{1}{M}$, then the identity \refE{alpha-quasiperfect-2} follows.
\end{IEEEproof}

Theorem \ref{thm:alpha-quasiperfect} shows that, whenever they exist, quasi-perfect codes attain the meta-converse bound with equality. Particularizing this result in the classical case we obtain \cite[Th.~1]{tit19qp}, which shows the optimality of the quasi-perfect binary codes for the BSC and  MDS codes for erasure channels. Definition~\ref{def:quasiperfect} extends the notion of generalized perfect and quasi-perfect codes to classical-quantum symmetric channels and Theorem \ref{thm:alpha-quasiperfect} shows their optimality.

In the classical setting the codes belonging to this class are rare and only exist for short blocklengths. Then, one may wonder if they exist at all for classical-quantum channels of interest. In the next section we show that this is the case for a family of $2$-qubit classical-quantum channels and certain code parameters.

%
%


\section{2-Qubit Classical-Quantum Channels and Bell Codes}\label{sec:bell-codes}

\subsection{Pure 2-qubit classical-quantum channel}

We consider a 2-qubit pure-state channel with output
\begin{align}\label{eqn:2qubit}
  \ket{\varphi} &\equiv \sum_{l=0}^{3}\alpha_l\ket{l}=\sum_{l=0}^{3}\alpha_l\ket{l_{1}l_0}=\alpha_0\ket{00} + \alpha_1\ket{01}+ \alpha_{2}\ket{10}+ \alpha_{3}\ket{11},
\end{align}
for $\sum_{l=0}^{3} |\alpha_l|^2=1$ and where $l_1 l_0$ are the binary digit representation of $l$. The channel is then defined by $\{W_x=\ket{\varphi_x}\bra{\varphi_x}\}$
We define the codebook $\Cc = \{x_1,\ldots,x_M\}$, $M = 2K \geq 4$,
such that the channel output is given by
\begin{align}
  \ket{\varphi_{x_m}} =
  \begin{cases}
    \frac{1}{\sqrt{2}}\bigl(\ket{00}+e^{j\phi_k}\ket{11}\bigr),& m=1+2k,\\
    \frac{1}{\sqrt{2}}\bigl(\ket{01}+e^{j\phi_k}\ket{10}\bigr),& m=2+2k,
  \end{cases}\label{eqn:bcode}
\end{align}
where $\phi_k = 2\pi k/K$, and $k=0\ldots K-1$.

For $M=4$, the channel outputs $\ket{\varphi_{x_m}}$ correspond precisely
to the Bell states~\cite{nielsen_chuang_2010}. For $M \geq 4$,
we refer to this family of codes as \textit{Bell codes},
since they follow from a generalization of the Bell states.

Since $\sum_{m=1}^M \ket{\varphi_{x_m}}\bra{\varphi_{x_m}} \geq \openone$
for $M \geq 4$, these codes are either perfect (when $M=4$) or quasi-perfect
(when $M>4$) for the 2-qubit pure-state channel, as shown in the example from
\refS{qp-codes}.

\begin{proposition}\label{prop:bcode}
The 2-quit classical-quantum channel $W_{x} = \ket{\varphi_{x}} \bra{\varphi_{x}}$ is
symmetric with respect to $\mu_0 = \frac{1}{4}\openone_4$ and the Bell code $\Cc$ is
quasi-perfect for this channel. Moreover, 
\begin{align}
  \Pe(\Cc) &= \alpha_{\frac{1}{M}} \bigl(W_x \,\|\, \mu_0\bigr)
    = 1 - \frac{4}{M}.
\end{align}
\end{proposition}
\begin{IEEEproof}
See Example 2 in \refS{qp-codes}, with error probability \refE{error-prob-pure-n}.
\end{IEEEproof}

When $M=4$, the code corresponds precisely to the (orthogonal) Bell states and the transmitted message can be determined without errors. For $M = 2P > 4$, the codewords are no longer orthogonal to each other and therefore they incur in measurement errors even for the ideal pure 2-qubit classical-quantum channel. Nevertheless, as shown in Theorem~\ref{thm:meta-converse} and in Proposition~\ref{prop:bcode}, there exist no other packing of pure states with lower error probability. This is not only true for the ideal channel but also when the transmission is affected by certain errors, as we will see now for the depolarizing and erasure channels.

\subsection{Classical-quantum depolarizing channel}

Consider the 2-qubit classical-quantum channel in (\ref{eqn:2qubit}) observed after a quantum depolarizing channel, defined~as
\begin{align}\label{eqn:depolarizing-channel}
\mathcal N^D_{A\rightarrow B}(\rho_A)=p\frac{1}{4}\openone_4+(1-p)\rho_A,
\end{align}

The combined classical-quantum channel is thus $W_x=\mathcal N^D_{A\rightarrow B}\bigl(\ket{\varphi_x}\bra{\varphi_x}_A\bigr)$. Using the Bell code defined in (\ref{eqn:bcode}), the channel output is given by $W_{m}=\mathcal N^D_{A\rightarrow B}\bigl(\ket{\varphi_{x_m}}\bra{\varphi_{x_m}}_A\bigr)$, $m=1,\ldots,M$.

\begin{proposition}\label{prop:bellDepolarizing}
Let $\mu_0 =\frac{1}{4}\openone_4$. Then, the 2-qubit classical-quantum depolarizing channel is symmetric with respect to $\mu_0$ and the Bell code $\Cc$ is quasi-perfect for this channel. Moreover, 
\begin{align}
  \Pe(\Cc) &= \alpha_{\frac{1}{M}} \bigl(W_x \,\|\, \mu_0\bigr)
    = 1-\frac{1}{M}(4-3p),
\end{align}
which is obtained using decoder $\Pc = \{ \Pi_{1},  \ldots, \Pi_{M} \}$ with $\Pi_{i}$ given by
\begin{align}\label{eqn:decoderDepolarizing}
  \Pi_{i}=\frac{4}{M}\ket{\varphi_{x_i}}\bra{\varphi_{x_i}}.
\end{align}
\end{proposition}
\begin{IEEEproof}
Consider the decoder $\Pc = \{ \Pi_{1},  \ldots, \Pi_{M} \}$ with $\Pi_{i}$ defined in  \refE{decoderDepolarizing}.
\subsubsection{Decoder optimality}
One can check that $\Pi_{i}\geq 0$ and $\sum_{i=1}^M \Pi_{i}= \openone_4$. For this decoder,
\begin{align}
       \Lambda(\Pc) &\triangleq \frac{1}{M} \sum_{i=1}^{M}  W_i \Pi_i =\frac{4}{M^2} \sum_{i=1}^{M}  W_i \ket{\varphi_{x_i}}\bra{\varphi_{x_i}}\\
	   &=\frac{1}{4M}(4-3p)\openone_4.
\label{eqn:LambdaPcD}
\end{align}
Then, it follows that 
\begin{align}
        \Lambda(\Pc)\Pi_i=\frac{1}{M}W_i\Pi_i,
\end{align}
which implies (\ref{eqn:mht-pistar1}). Equation (\ref{eqn:mht-pistar2}) is satisfied since, for  arbitrary unit norm vector $\ket{\psi}$,
\begin{align}
       \frac{\bra{\psi}\Lambda(\Pc)\ket{\psi}}{\frac{1}{M}\bra{\psi}W_i\ket{\psi}}&=
       \frac{\frac{1}{4M}(4-3p)}{\frac{1}{4M}(p+4(1-p)|\braket{\psi|\varphi_{x_i}}|^2 )}\geq
       \frac{4-3p}{p+4(1-p)}=1\label{eqn:cauchy-schwarzD}
\end{align}
So $\Pc = \{ \Pi_{1},  \ldots, \Pi_{M} \}$ minimizes the error probability for the Bell code $\Cc$.

\subsubsection{Symmetry of the channel with respect to $\mu_0$}
We will prove next that
\begin{align}\label{eqn:Sx_umbrellaD}
\Ec_{x}(t,\mu_0)
= \begin{cases}
     \openone_4, & t < 0,\\
     \ket{v}\bra{v}, & 0 \leq t \leq t_{0},\\
     0, & t > t_{0},
  \end{cases}
\end{align}
for $\ket{v} = \ket{\varphi_x}$ and $t_0=4-3p$ independent of $x$.
Then, using \refE{Sx_umbrellaD} in $F_{x}(t,\mu_0) = \tr\bigl( W_x \Ec_{x}(t,\mu_0) \bigr)$, it yields
\begin{align}
F_{x}(t,\mu_0)
= \begin{cases}
     1, & t  < 0,\\
     1-\frac{3}{4}p, & 0 \leq t \leq t_{0},\\
     0, & t > t_{0},
  \end{cases}
\end{align}
and $\tr\bigl(W_x \ket{v}\bra{v})$ is independent of $\varphi_x$, so the channel is symmetric with respect to $\mu_0$.

It remains to show that \refE{Sx_umbrellaD} holds. The identity for $t < 0$ follows trivially. We consider an arbitrary  unit-norm vector $\ket{v}$. Then, the largest eigenvalue of $W_x - t \mu_0$ is given by
\begin{align}
  \max_{v} \bra{v} \left(W_x
      - t \mu_0 \right) \ket{v}
&= \max_{v} \Bigl\{
   \frac{p}{4}+(1-p)|\braket{v|\varphi_{x}}|^2 - \frac{t}{4} \Bigr\}\\
&= 1-\frac{3}{4}p-\frac{t}{4}.\label{eqn:lambda_max_umbrellaD3}
\end{align}
The eigenvalue \refE{lambda_max_umbrellaD3} is negative for $t>4-3p$ and non-negative otherwise. Then, we obtain that $F_{x}(t,\mu_0) = 0$, for $t > 4-3p$. For $0 \leq t \leq 4-3p$, \refE{lambda_max_umbrellaD3} is the only non-negative eigenvalue with associated eigenvector $\ket{v} = \ket{\varphi_x}$. Therefore, considering the three regions, we obtain $\refE{Sx_umbrellaD}$.

\subsubsection{$\Cc$ is quasi-perfect with respect to $\mu_0$}
Comparing \refE{LambdaPcD} with the auxiliary state $\mu_0$ considered in the statement of Proposition \ref{prop:bellDepolarizing}, we observe that
\begin{align}
  \mu_0 = \frac{1}{c_0} \Lambda(\Pc) =  \frac{1}{M c_0} \sum_{m=1}^{M}  W_m \Pi_m,
\end{align}
where $c_0=\frac{4-3p}{M}$ is a normalizing constant and where $\Pc$ satisfies the optimality conditions.

Take $t={M c_0}=4-3p$, then $\frac{1}{M}W_m -\Lambda(\Pc)$ is negative semidefine and $\Sphi{x_m}(t,\mu_0)=0$. As a result,  $\bigl\{\Sphi{x_m}(t,\mu_0) \bigr\}_{x \in \Cc}$ are orthogonal to each other.
Similarly, for this choice of $t$ and $\mu_0$, it follows that $\Ec_{x_m}(t,\mu_0)=\ket{\varphi_{x_m}}\bra{\varphi_{x_m}}$. Therefore $\sum_{x\in\Cc} \Ec_{x}(t,\mu) =\frac{M}{4} \openone_4 \geq \openone_4$ and the code is quasi-perfect.

\subsubsection{Error probability}
Using Theorem \ref{thm:alpha-quasiperfect}, it follows that $\Pe(\Cc) = \alpha_{\frac{1}{M}} \bigl(W_x \,\|\, \mu_0\bigr)$.
Moreover, using the optimal decoder $\Pc$, we obtain
\begin{align}
  \Pe(\Cc) &= 1 - \frac{1}{M} \sum_{i=1}^M  \tr\bigl(W_i \Pi_{i}\bigr)\\
           &= 1 - \tr\bigl(\Lambda(\Pc)\bigr)\\
           &= 1 - \frac{4-3p}{M},
\end{align}
where in the last step we used \refE{LambdaPcD}.
\end{IEEEproof}

\subsection{Classical-quantum erasure channel}

We consider the classical-quantum channel (\ref{eqn:2qubit}) observed after a quantum erasure channel, defined as
\begin{align}\label{eqn:erasure-channel}
\mathcal N^E_{A\rightarrow B}(\rho_A)=(1-\epsilon)\mathcal{I}_{A\rightarrow B}(\rho_A) +\epsilon |e \rangle\langle e|_B.
\end{align}
where the Isometric channel $\mathcal{I}_{A\rightarrow B}(\rho_A)=I_{A\rightarrow B}\rho_A I_{A\rightarrow B}^{\dagger}$ is defined using the isometry
\begin{align}
I_{A\rightarrow B}=\begin{bmatrix} \openone_4\\\begin{matrix} 0 & \hdots & 0 \end{matrix} \end{bmatrix}
\end{align}
 as unique Kraus operator and where $\{|00\rangle, |01\rangle,|10\rangle, |11\rangle,|e\rangle\}$ form an orthonormal basis in $\mathcal H_B$. The combined classical-quantum channel is then $W_x = \mathcal N^E_{A\rightarrow B}\bigl( \ket{\varphi_x}\bra{\varphi_x}_A\bigr)$.
 

\begin{proposition} \label{prop:bellErasure}
Let
\begin{align}
\mu_0 = 
\frac{1}{4-3\epsilon}\begin{bmatrix} (1-\epsilon)\openone_4 & \begin{matrix} 0\\ \vdots \\ 0 \end{matrix} \\ \begin{matrix} 0 & \cdots & 0 \end{matrix} & \epsilon \end{bmatrix}.
\end{align}
Then, the 2-quit classical-quantum erasure channel is symmetric with respect to $\mu_0$ and the Bell code $\Cc$ is quasi-perfect for this channel. Moreover, 
\begin{align}
  \Pe(\Cc) &= \alpha_{\frac{1}{M}} \bigl(W_x \,\|\, \mu_0\bigr)
    = 1-\frac{1}{M}(4-3\epsilon).
\end{align}
\end{proposition}
\begin{IEEEproof}
For $M\geq 4$, the channel output $\rho_B$ induced by the code $\Cc$ is given by
\begin{align}
  \rho_B&= \frac{1}{M}\sum_{m=1}^M W_m=\frac{1}{M}\sum_{m=1}^M \mathcal N_{A\rightarrow B}\bigl(\ket{\varphi_{x_m}}\bra{\varphi_{x_m}}_A\bigr)= 
\begin{bmatrix} (1-\epsilon)\frac{1}{4}\openone_4 & \begin{matrix} 0\\ \vdots \\ 0 \end{matrix} \\ \begin{matrix} 0 & \hdots & 0 \end{matrix} & \epsilon \end{bmatrix}.
\end{align}
We define the decoder $\Pc = \{ \Pi_{1},  \ldots, \Pi_{M} \}$ as
\begin{align}
  \Pi_{m}=\frac{1}{M}\rho_B^{-\frac{1}{2}} W_m\rho_B^{-\frac{1}{2}}=\frac{1}{M} \begin{bmatrix} 4\ket{\varphi_{x_m}}\bra{\varphi_{x_m}} & \begin{matrix} 0 \\ \vdots \\0 \end{matrix} \\ \begin{matrix} 0 &\hdots  & 0 \end{matrix} & 1 \end{bmatrix}.
\end{align}
\subsubsection{Decoder optimality}
It can be veried that $\Pi_{m}\geq 0$ and that $\sum_{m=1}^M \Pi_{m}= \openone_5$, and
\begin{align}
       \Lambda(\Pc) = \frac{1}{M} \sum_{m=1}^{M}  W_m \Pi_m
	   =\frac{1}{M}\begin{bmatrix} (1-\epsilon)\openone_4 & \begin{matrix} 0 \\ \vdots \\ 0 \end{matrix} \\ \begin{matrix} 0 &\hdots  & 0 \end{matrix} & \epsilon \end{bmatrix}.
\label{eqn:LambdaPcE}
\end{align}

Since $\Lambda(\Pc)\Pi_m=\frac{1}{M}W_m\Pi_m$, the condition (\ref{eqn:mht-pistar1}) is satisfied. The condition (\ref{eqn:mht-pistar2}) is satisfied since, for an arbitrary unit norm vector $\ket{\psi'}\triangleq \bigl[\begin{smallmatrix} \ket{\psi} \\ \pi \end{smallmatrix}\bigr]$, where $|\pi|\leq 1$, $\braket{\psi|\psi}=1-|\pi|^2$,
\begin{align}
       \frac{\bra{\psi'}\Lambda(\Pc)\ket{\psi'}}{\frac{1}{M}\bra{\psi'}W_m\ket{\psi'}}&=
       \frac{\frac{1}{M}\left[(1-\epsilon)\braket{\psi|\psi}+\epsilon  |\pi|^2 \right]}{\frac{1}{M}\left[(1-\epsilon)|\braket{\psi|\varphi_{x_m}}|^2+\epsilon  |\pi|^2 \right]}
       \geq 1,\label{eqn:cauchy-schwarzE}
\end{align}
Since $|\braket{\psi|\varphi_{x_m}}|^2\leq \braket{\psi|\psi}\braket{\varphi_{x_m}|\varphi_{x_m}}=\braket{\psi|\psi}$ and \refE{mht-pistar2} follows.
We conclude that $\Pc = \{ \Pi_{1},  \ldots, \Pi_{M} \}$ minimizes the error probability for the Bell code $\Cc$.

\subsubsection{Symmetry of the channel with respect to $\mu_0$}
We study the eigenvalues of $\Ec_{x}(t,\mu_0) = \bigl\{W_x-t\mu_0 \geq 0\bigr\}$. 
First, for $t < 0$, $\Ec_{x}(t,\mu_0)=\openone_5$ holds trivially since both $W_x\geq 0$
and $\mu_0 \geq 0$. Then, for $t < 0$, $F_{x}(t,\mu_0)=1$.

For $t \geq 0$, we write
\begin{align}
W_x-t\mu_0 &= \begin{bmatrix}
(1-\epsilon)\ket{\varphi_x}\bra{\varphi_x}  & 0 \\
 0   &  \epsilon \\ \end{bmatrix} - \frac{t}{4-3\epsilon}
\begin{bmatrix} (1-\epsilon)\openone_4  & 0 \\
 0   &  \epsilon \\  
\end{bmatrix} \\ &= \begin{bmatrix}
(1-\epsilon)\bigl(\ket{\varphi_x}\bra{\varphi_x}-\frac{t}{4-3\epsilon} \openone_4\bigr)  & 0 \\
 0   &  \epsilon\bigl(1-\frac{t}{4-3\epsilon}\bigr) \\ \end{bmatrix}
\end{align}
For $t>4-3\epsilon$ the matrix $W_x-t\mu_0$ has no positive eigenvalues and therefore $F_{x}(t,\mu_0)=0$. For $0 \leq t \leq 4-3\epsilon$ it has two positive eigenvalues whose eigenvectors are $\ket{v'_1}=\varphi_x$ and $\ket{v'_2}=\bigl[0, 0, 0, 0, 1\bigr]^{T}$. In this case:
\begin{align}
F_{x}(t,\mu_0) &=  \tr\bigl (W_x (\ket{v'_1}\bra{v'_1}+\ket{v'_2}\bra{v'_2})\bigr) \\ &= \tr\left (\begin{bmatrix} 
(1-\epsilon)\ket{\varphi_x}\bra{\varphi_x}  & 0 \\
 0   &  \epsilon \\ \end{bmatrix} \begin{bmatrix} 
\ket{\varphi_x}\bra{\varphi_x}  & 0 \\
 0   &  1 \\ \end{bmatrix}\right) \\ &= (1-\epsilon) + \epsilon = 1.
\end{align}
We conclude that 
\begin{align}\label{eqn:Fx_umbrellaE}
F_{x}(t,\mu_0)
= \begin{cases}
     1, & t  \leq 4-3\epsilon\\
     0, & t > 4-3\epsilon,
  \end{cases}
\end{align}
which does not depend on the value of $x$. Then, the channel is symmetric with with respect to $\mu_0$.

\subsubsection{$\Cc$ is quasi-perfect with respect to $\mu_0$}

Comparing \refE{LambdaPcE} with $\mu_0$ defined in the statement of Proposition \ref{prop:bellErasure}, we conclude that
\begin{align}
  \mu_0 = \frac{1}{c_0} \Lambda(\Pc) =  \frac{1}{M c_0} \sum_{m=1}^{M}  W_m \Pi_m,
\end{align}
where $c_0=\frac{4-3\epsilon}{M}$ and where $\Pc$ satisfies the optimality conditions. 

Take $t={M c_0}=4-3\epsilon$,  then $\frac{1}{M}W_m -\Lambda(\Pc)$ is negative semidefine. Hence, $\Ec_{x_m}(t,\mu_0) =\Spho{x_m}(t,\mu_0)$ and $\Sphi{x_m}(t,\mu_0)=0$, so $\bigl\{  \Ec_{x}^\bullet(t,\mu) \bigr\}_{x \in \Cc}$ are orthogonal to each other.
For this choice of $t$ and $\mu_0$,
\begin{align}
       \Ec_{x_m}(t,\mu_0^*)&=\begin{bmatrix} \ket{\varphi_x}\bra{\varphi_x}  & 0 \\ 0   &  1 \\ \end{bmatrix},
\end{align}
we conclude that $\sum_{x\in\Cc} \Ec_{x}(t,\mu) =\begin{bmatrix} \frac{M}{4}\openone_4  & 0 \\ 0   &  M \\ \end{bmatrix} \geq \openone_5$, which means that the code is quasi-perfect.

\subsubsection{Error probability}
From Theorem \ref{thm:alpha-quasiperfect}, it follows that $\Pe(\Cc) = \alpha_{\frac{1}{M}} \bigl(W_x \,\|\, \mu_0\bigr)$.
Moreover, using the optimal decoder $\Pc$, we obtain
\begin{align}
  \Pe(\Cc) &= 1 - \frac{1}{M} \sum_{m=1}^M  \tr\bigl(W_m \Pi_{m}\bigr)\\
           &= 1 - \tr\bigl(\Lambda(\Pc)\bigr)\\
           &= 1 - \frac{4-3\epsilon}{M},
\end{align}
where in the last step we used \refE{LambdaPcE}.
\end{IEEEproof}

\subsection{Extension to $N$-qubit classical-quantum channels}

Consider now an arbitrary  $N$-qubit classical-quantum channel with pure outputs given by
\begin{align}\label{eqn:N-qubit-ideal-channel}
  \ket{\varphi} &\equiv \sum_{l=0}^{2^N-1}\alpha_l\ket{l}=\sum_{l=0}^{2^N-1}\alpha_l\ket{l_{N-1} \ldots l_0}=\alpha_0\ket{0\ldots00} + \alpha_1\ket{0\ldots 01}+ \alpha_{2^N-1}\ket{1\ldots 11}
\end{align}
for $\sum_{l=0}^{2^N-1} |\alpha_l|^2=1$ and where $l_{N-1} \dots l_0$ are the digits of the binary representation of $l$. The channel is then given by $\{W_x=\ket{\varphi_x}\bra{\varphi_x}\}$.
For $M=2^{N-1}K\geq 2^{N}$, we define the $N$-qubit Bell codebook of cardinality $M$ given by $\Cc = \bigl\{ x_{1},\ldots, x_{M}\bigr\}$ with channel outputs
\begin{align}
  \ket{\varphi_{x_m}} =
  \begin{cases}
    \frac{1}{\sqrt{2}}\bigl(\ket{00}+e^{j\phi_k}\ket{11}\bigr)\otimes \ket{l_{N-3} \dots l_0},& m=1+2k+2Kl,\\
    \frac{1}{\sqrt{2}}\bigl(\ket{01}+e^{j\phi_k}\ket{10}\bigr)\otimes \ket{l_{N-3} \dots l_0},& m=2+2k+2Kl,
  \end{cases}\label{eqn:bNcode}
\end{align}
where $\phi_k = 2\pi k/K$, $k=0,\ldots, K-1$, and $l=0,\ldots,2^{N-2}-1$.

 The channel ouput for codeword $x_m$ is thus given by the pure state $W_{m}=\ket{\varphi_{x_m}}\bra{\varphi_{x_m}}$.

\begin{proposition}\label{prop:bellNqubit}
Let $\mu_0 = \frac{1}{2^N}\openone_{2^N}$. The $N$-quit classical-quantum channel is symmetric with respect to $\mu_0$ and the $N$-qubit Bell code $\Cc$ is quasi-perfect for this channel. Moreover, 
\begin{align}
  \Pe(\Cc) &= \alpha_{\frac{1}{M}} \bigl(W_x \,\|\, \mu_0\bigr)
    = 1 - \frac{2^N}{M}.
\end{align}
which is obtained using decoder $\Pc = \{ \Pi_{1},  \ldots, \Pi_{M} \}$ with
\begin{align}
  \Pi_{i}=\frac{2^N}{M} W_i.
\end{align}
\end{proposition}

\begin{IEEEproof}
This result is a generalization of Proposition~\ref{prop:bcode} and the proof follows similar steps.
\end{IEEEproof}

The $N$-qubit Bell code is also quasi-perfect for channels affected by depolarization 
and erasures, as stated by the following results which are the analogous to Propositions~\ref{prop:bellDepolarizing} and~\ref{prop:bellErasure}. 

Consider the $N$-qubit classical-quantum channel in (\ref{eqn:N-qubit-ideal-channel}) observed after a quantum depolarizing channel:
\begin{align}
\mathcal N^D_{A\rightarrow B}(\rho_A)=p\frac{1}{2^N}\openone_{2^N}+(1-p)\rho_A,
\end{align}

The combined classical-quantum channel is thus $W_x=\mathcal N^D_{A\rightarrow B}\bigl(\ket{\varphi_x}\bra{\varphi_x}_A\bigr)$. Using the Bell code defined in (\ref{eqn:bNcode}), the channel output is given by $W_{m}=\mathcal N^D_{A\rightarrow B}\bigl(\ket{\varphi_{x_m}}\bra{\varphi_{x_m}}_A\bigr)$, $m=1,\ldots,M$.
\begin{proposition}\label{prop:bellDepolarizingN}
Let $\mu_0 =\frac{1}{2^N}\openone_{2^N}$. Then, the $N$-qubit classical-quantum depolarizing channel is symmetric with respect to $\mu_0$ and
the $N$-qubit Bell code $\Cc$ is quasi-perfect for this channel. Moreover, 
\begin{align}
  \Pe(\Cc) &= \alpha_{\frac{1}{M}} \bigl(W_x \,\|\, \mu_0\bigr)
    = 1-\frac{1}{M}(2^N(1-p)+p).
\end{align}
which is obtained using decoder $\Pc = \{ \Pi_{1},  \ldots, \Pi_{M} \}$ with
\begin{align}
  \Pi_{i}=\frac{2^N}{M}\ket{\varphi_{x_i}}\bra{\varphi_{x_i}}.
\end{align}
\end{proposition}
\begin{IEEEproof}
This result is a generalization of Proposition~\ref{prop:bellDepolarizing} and the proof follows similar steps.
\end{IEEEproof}

Finally consider the classical-quantum channel (\ref{eqn:N-qubit-ideal-channel}) observed after a quantum erasure channel, defined as
\begin{align*}
\mathcal N^E_{A\rightarrow B}(\rho_A)=(1-\epsilon)\mathcal{I}_{A\rightarrow B}(\rho_A) +\epsilon |e \rangle\langle e|_B.
\end{align*}
where the Isometric channel $\mathcal{I}_{A\rightarrow B}(\rho_A)=I_{A\rightarrow B}\rho_A I_{A\rightarrow B}^{\dagger}$ is defined using the isometry
\begin{align}
I_{A\rightarrow B}=\begin{bmatrix} \openone_{2^N}\\\begin{matrix} 0 & \hdots & 0 \end{matrix} \end{bmatrix}
\end{align}
 as unique Kraus operator and where $\{|0\rangle, \ldots,|2^N-1\rangle,|e\rangle\}$ form an orthonormal basis in $\mathcal H_B$. The combined classical-quantum channel is then $W_x = \mathcal N^E_{A\rightarrow B}\bigl( \ket{\varphi_x}\bra{\varphi_x}_A\bigr)$.
 
\begin{proposition}\label{prop:bellErasureN}
Let 
\begin{align}
\mu_0 =\frac{1}{2^N(1-\epsilon)+\epsilon}\begin{bmatrix} (1-\epsilon)\openone_{2^N} & \begin{matrix} 0\\ \vdots \\ 0 \end{matrix} \\ \begin{matrix} 0 & \cdots & 0 \end{matrix} & \epsilon \end{bmatrix}.
\end{align}
Then, the $N$-qubit classical-quantum erasure channel is symmetric with respect to $\mu_0$ and the $N$-qubit Bell code $\Cc$ is quasi-perfect for this channel. Moreover, 
\begin{align}
  \Pe(\Cc) &= \alpha_{\frac{1}{M}} \bigl(W_x \,\|\, \mu_0\bigr)
    = 1-\frac{1}{M}(2^N(1-\epsilon)+\epsilon).
\end{align}
which is obtained using decoder $\Pc = \{ \Pi_{1},  \ldots, \Pi_{M} \}$ with
\begin{align}
  \Pi_{i}=\frac{1}{M} \begin{bmatrix} 2^N\ket{\varphi_{x_m}}\bra{\varphi_{x_m}} & \begin{matrix} 0 \\ \vdots \\0 \end{matrix} \\ \begin{matrix} 0 &\hdots  & 0 \end{matrix} & 1 \end{bmatrix}.
\end{align}
\end{proposition}
\begin{IEEEproof}
This result is a generalization of Proposition~\ref{prop:bellErasure} and the proof follows similar steps.
\end{IEEEproof}

\section{Discussion}\label{sec:discussion}

In this work we explored the connections between hypothesis testing and classical-quantum channel coding.
First, we obtained two alternative exact expressions for the minimum error probability of multiple quantum hypothesis testing when a (classical) prior distribution is placed over the hypotheses. The expression in Theorem~\ref{thm:main-result} illustrates connections among the different settings of hypothesis testing and Corollary~\ref{cor:tight-spectrum} provides an alternative formulation based on information-spectrum measures.
A direct application of these results to a classical-quantum channel coding problem shows that Matthews-Wehner converse bound \cite[Th.~19]{matthews2014finite} and Hayashi-Nagaoka lemma\mbox{\cite[Lemma~4]{hayashi2003general}} with certain parameters yield the exact error probability in this setting. 

While these results are of theoretical interest, the resulting expressions still depend on the the codebook and their application as performance benchmarks for classical-quantum channels is limited. We studied different relaxations and connections with practical converse bounds in the literature, thus characterizing the weaknesses of these bounds and the gap to the exact channel-coding error probability.
Of special interest for this work is the so-called meta-converse bound~\cite[Eq.~(46)]{matthews2014finite}, presented here in Theorem~\ref{thm:meta-converse}, which corresponds to the error probability of a binary hypothesis test with certain parameters.

In the second part of this work, we introduced the notion of perfect and quasi-perfect codes for symmetric classical-quantum channels. It is interesting to note that this notion is channel dependent --since a code being perfect for a channel it is not necessarily perfect for another one-- and that it encompasses classical perfect and quasi-perfect codes as a special case~\cite[Sec.~IV]{tit19qp}. Therefore, this definition includes as special cases perfect and quasi-perfect binary codes for the BSC and MDS codes for classical erasure channels. Theorem~\ref{thm:Pe-quasiperfect} provides an expression of the error probability of perfect and quasi-perfect codes for symmetric classical-quantum channels, which is then used in Theorem~\ref{thm:alpha-quasiperfect} to prove that these codes attain the meta-converse bound with equality. These codes, whenever they exist, are thus optimal in the sense that they achieve the smallest error probability among all codes of the same blocklength and cardinality.

Establishing the existence of generalized perfect and quasi-perfect codes for a given set of system parameters is a difficult problem, even for simple classical channels. For instance, \cite{etzion2005} studies their existence for the BSC channel and~\cite{Seroussi1986} shows that MDS codes, which are generalized quasi-perfect for the $q$-ary erasure channel, only exist for blocklengths $n\leq q+1$.
In this work, we consider a family of 2-qubit classical-quantum channels affected by quantum erasures or by depolarization. Using the framework presented, we established that a generalization of Bell states, that we name Bell codes, are quasi-perfect for these channels when their cardinality is $M\geq 4$. For these channels and code parameters, we have thus established the error probability and structure of the best coding scheme. Proving the existence of perfect and quasi-perfect codes for other classical-quantum channels of practical interest is an unexplored line of research.


\bibliographystyle{IEEEtran}
\bibliography{bib/references}

\end{document}